%% file: ms.tex
\documentclass[sigconf]{acmart}

\usepackage{booktabs} 
\usepackage{url}
\usepackage{graphicx}
\usepackage{subfigure}
\usepackage{soul}
\usepackage[draft,inline,nomargin]{fixme}

\usepackage{xspace}
\usepackage{algorithm}
\usepackage{algorithmicx}
\usepackage{algpseudocode}

\usepackage{todonotes}
\usepackage{flushend}

\begin{document}
\title{POISED: Spotting Twitter Spam Off the Beaten Paths}

\author{Shirin Nilizadeh}
\affiliation{%
  \institution{UC Santa Barbara}
  \state{California}
  \country{USA}
}
\email{nilizadeh@ucsb.edu}

\author{Fran\c{c}ois Labr\`eche}
\affiliation{%
  \institution{Ecole Polytechnique de Montr\'eal}
  \state{Qu\'ebec}
  \country{Canada}
}
\email{francois.labreche@polymtl.ca}

\author{Alireza Sedighian}
\affiliation{%
  \institution{Ecole Polytechnique de Montr\'eal}
  \state{Qu\'ebec}
  \country{Canada}
}
\email{alireza.sadighian@polymtl.ca}

\author{Ali Zand}
\affiliation{%
  \institution{UC Santa Barbara}
  \state{California}
  \country{USA}
}
\email{zand@ucsb.edu}

\author{Jos\'e Fernandez} 
\affiliation{%
 \institution{Ecole Polytechnique de Montr\'eal}
  \state{Qu\'ebec}
  \country{Canada}
  }
\email{jose.fernandez@polymtl.ca}
  
\author{Christopher Kruegel}
\affiliation{%
  \institution{UC Santa Barbara}
  \state{California}
  \country{USA}
}
\email{chris@ucsb.edu}

\author{Gianluca Stringhini}
\affiliation{%
  \institution{University College London}
  \city{London}
  \country{United Kingdom}
  }
\email{g.stringhini@ucl.ac.uk}

\author{Giovanni Vigna}
\affiliation{
\institution{UC Santa Barbara}
  \state{California}
  \country{USA}
}
\email{vigna@ucsb.edu}

%

\def\approach{{\sc POISED}\xspace}

\begin{abstract}
Cybercriminals have found in online social networks a propitious 
medium to spread spam and malicious content. 
Existing techniques for detecting spam include 
predicting the trustworthiness of accounts 
and analyzing the content of these messages. 
However, advanced attackers can still successfully evade these defenses. 

Online social networks bring people who have personal connections or share common interests to form communities. In this paper, we first show that users within a networked community share some topics of interest. 
Moreover, content shared on these social network tend to propagate according to the interests of people. Dissemination paths may emerge where some communities post similar messages, based on the interests of those communities. 
Spam and other malicious content, on the other hand, follow different spreading patterns.
 
In this paper, we follow this insight and present \approach, a system that leverages the differences in propagation between benign and malicious messages on social networks to identify spam and other unwanted content. We test our system on a dataset of 1.3M tweets collected from 64K users, and we show that our approach is effective in detecting malicious messages, reaching 91\% precision and 93\% recall. 
We also show that \approach's detection is more comprehensive than previous systems,
by comparing it to three state-of-the-art spam detection systems that have been proposed by the research community in the past. \approach significantly outperforms each of these systems. 
Moreover, through simulations, we show how \approach is effective in the early detection of spam messages and how it is resilient against two well-known adversarial machine learning attacks.
\end{abstract}

%
%


\copyrightyear{2017}
\acmYear{2017}
\setcopyright{acmcopyright}
\acmConference{CCS '17}{October 30-November 3, 2017}{Dallas, TX, USA}\acmPrice{15.00}\acmDOI{10.1145/3133956.3134055}
\acmISBN{978-1-4503-4946-8/17/10}

\fancyhead{}
\settopmatter{printacmref=false, printfolios=false}


\maketitle

\input{introduction}
\input{model}

\input{method}

\input{eval}

\input{results-h1}
\input{results-h2}
\input{related-work}
\input{discussion}

\input{conclusion}
\section*{Acknowledgements}

This research was supported by the EPSRC under Grant N008448 and by the European Commission as part of the ENCASE project (H2020-MSCA RISE of the European Union under GA number 691025). It is also in part supported by ESET and FRQNT.

This material is also based upon work supported by the National Science Foundation under Awards No. DGE-1623246, and CNS-1408632, as well as by Office of Naval Research under Award No. N00014-17-1-2011, and in part is supported by Google. 
Any opinions, findings, and conclusions or recommendations expressed in this publication are those of the authors and do not necessarily reflect the views of the above agencies. Also, this material is based on research sponsored by DARPA under agreement number FA8750-15-2-0084. The U.S. Government is authorized to reproduce and distribute reprints for Governmental purposes notwithstanding any copyright notation thereon.
The views and conclusions contained herein are those of the authors and should not be interpreted as necessarily representing the official policies or endorsements, either expressed or implied, of DARPA or the U.S. Government.

\bibliographystyle{ACM-Reference-Format}
\bibliography{refs} 

\end{document}

%% file: introduction.tex
\section{Introduction}
Cybercriminals have found in social networks a propitious medium to spread malicious content and perform scams against users~\cite{grier2010spam}. 
Social networks are leveraged by cybercriminals for a number of reasons. 
First, social networks are very popular, with the largest ones having hundreds of millions of users: this constitutes a large victim base for criminals. 
Second, attackers who compromise social network accounts with an already established reputation can exploit the inherent trust between connected users to spread malicious content very effectively~\cite{egele2013compa, jagatic2007social}.

Previous work addressed the detection of spam on social networks by predicting the trustworthiness of the accounts that post messages, notably by detecting Sybil communities~\cite{wang2013you, danezis2009sybilinfer}, bots~\cite{ferrara2014rise, stringhini2010detecting,wang2010detecting}, compromised
accounts~\cite{egele2013compa}, or a combination of these~\cite{viswanath2014towards, cao2014uncovering}.
Recent research, however, showed how attackers can successfully evade both Sybil-based defenses~\cite{liu2015exploiting} and account-based ones~\cite{yang2011free}. 
This happens because existing spam detection systems detect \emph{the way in which malicious accounts infiltrate the network and build connections}, rather than \emph{the way in which malicious messages spread across the network in comparison to legitimate ones}.

In this paper, we propose a novel way to detect malicious messages on social networks. Instead of looking at the characteristics of accounts or messages, we inspect the way in which messages spread on the social network. 

In social networks, users tend to form \emph{networked communities},  where most users are connected to many other users within the same community. These communities can be recognized by their structure in the underlying connection graph, as they form strongly connected subgraphs. 
For this reason, they are also dubbed \emph{structural communities}. The reasons why such communities form are as varied as the reasons why people connect to each other, such as family, geographical location, past common history, etc. Nonetheless, it has been recognized that one important reason why members of social networks tend to connect to others is the so-called \emph{homophily principle}, \textit{i.e.,} people connect to other people who hold similar thoughts and values~\cite{mcpherson2001birds}. In that sense, these people also form \emph{communities of interest}, where connected users communicate and interact on topics of common interest. In principle, members of a networked community may not necessarily share the same interests, and thus, structural communities and communities of interest may not coincide. However, we postulate that the homophily principle constitutes indeed the principal reason why people connect. 
Therefore, we formalize our first hypothesis as follows:

\textbf{H1}: In social networks, the topics of interest of users within a networked community are strongly shared among its members. In other words, networked communities are structured subsets of the larger set of users interested in the same topics.

It is recognized that the topology of social networks has an extremely important role in the dissemination of information~\cite{weng2013virality, nematzadeh2014optimal}. 
The dissemination of information is shaped by the structure of the network, and in particular faster dissemination is favored within networked communities~\cite{lerman2010information}. 
Nonetheless, such dissemination can traverse networked communities as long as there are members in both communities who share the same interests. 
As time progresses, dissemination paths may emerge where some communities trigger and post specific messages based on the interests of those communities and of the surrounding ones. 
These dissemination paths can help us predict patterns of postings within and outside of communities. For example, if two communities $C_1$ and $C_2$ always post messages on similar topics, then when a message is observed in $C_1$, the same or similar message has a high probability to also be posted or shared in $C_2$. 
In this paper, we refer to these communities interested in the same set of topics as \emph{parties of interest}. 

On the other hand, spam typically spread differently throughout the network. For example, messages that are posted by compromised accounts may spread in unexpected communities, because each compromised user posts that message regardless of whether the topic is of interest to the account owner or of the communities of interest of which the compromised user is a member~\cite{egele2015towards}.
This leads to formulate our second hypothesis as follows:

\textbf{H2}: Normal messages disseminate through predictable \emph{parties of interest} that include intra-community communication and inter-community exchanges between structural communities that share common interests. Conversely, the propagation probability of malicious messages through these parties of interest do not match with those of normal messages.

In this paper, we investigate these two hypotheses through experimentation on the Twitter social network. 
First, our analysis shows that, on Twitter, community members have a similar and restricted set of topics of interest, thus validating our first hypothesis. 
We then build a system, called \approach{}, that is able to detect whether a message shared on a social network spreads 
through expected parties of interest, or if it rather spreads anomalously. Our experimental performance evaluation shows that \approach{} can detect malicious spam messages with high accuracy, thus validating our second hypothesis. 

In a nutshell, \approach{} works as follows. 
First, it detects networked communities in Twitter by partitioning its social graph. Second, it identifies topics of interest in these communities. 
Then, it tracks the dissemination of similar messages through communities and constructs a probabilistic model of the parties of interest through which these messages are normally disseminated. 
Finally, leveraging this model, a classifier detects malicious content by identifying the messages that do not follow these expected parties of interest. 
\approach{} can successfully detect spam messages with 91\% precision and 93\% recall. We also compare \approach{} to three state-of-the-art spam detection systems that have been proposed by previous work: 
\textsc{SpamDetector}~\cite{stringhini2010detecting}, \textsc{Compa}~\cite{egele2013compa}, and \textsc{BotOrNot}~\cite{davis2016botornot}. 
With respect to the F1-score, \approach{} outperformed them by more than 70\%, 35\%, and 83\%, accordingly.

Through simulations, we show that \approach{} performs very well in detecting spam messages early on. 
For example, it can detect spam messages that have spread through only 20\% of the communities with 88\% precision and 75\% recall. 

Finally, we investigate the resilience of \approach{} against two common adversarial machine learning attacks~\cite{laskov2010machine}, \emph{poisoning}~\cite{rubinstein2009antidote} and \emph{evasion} attacks~\cite{biggio2013evasion}. 
Our simulation results suggest that the adversary needs to have a
great knowledge about the network and parties of interest to highly impact the performance of \approach{}. 
For example, even if 30\% of the network is compromised, the precision and recall remain at 82\% and 87\%, in the case of a poisoning attack, and at 75\% and 52\%, in the case of an evasion attack.

In summary, this paper makes the following contributions:

\begin{enumerate}
 \item Through our experiments on 300 Twitter neighborhoods with more than 15M tweets and 82K users, we show that networked communities are built around shared topics.

 \item We developed \approach{}, which relies on a combination of techniques from network science, natural language processing, and machine learning to detect spam messages by predicting the dissemination of messages through parties of interest. We tested \approach{} on a ground-truth dataset including data for 202 neighborhoods in Twitter with about 1.3M tweets and 64K users. Our results suggest that our approach is successful in detecting spam messages. Moreover, it outperforms other state-of-the-art detection systems.  
 
\item Our results demonstrate that this approach is scalable and can detect spam with only a partial knowledge of the social network. 
Our simulation results show that spam messages can be detected early on, when only attaining 20\% of their potential reach in their neighborhood network. 
We also show that \approach{} is difficult to evade for an active adversary. We simulate two attacks in which the adversary attempts to mimic the propagation of benign messages, and show that \approach{} is still highly effective even when the adversary has compromised a large portion of the network. 

\end{enumerate}

%% file: model.tex
\section{Background and Threat Model}
\label{model}

\subsection{Threat Model}

In our threat model, spam messages are posted on a large scale~\cite{twitterspam,thomas2011suspended} and are similar in content and format since, in most cases, they are generated by similar templates~\cite{gao2010detecting, gao2012towards}. 
This can be accomplished either by creating fake (Sybil) accounts~\cite{danezis2009sybilinfer}, compromising and abusing legitimate accounts~\cite{egele2013compa}, or by purchasing bots~\cite{fleizach2007slicing}. 

Unlike other related work that focuses on analyzing message content (\textit{i.e.,}~URLs)~\cite{lee2012warningbird, thomas2011design}, or finding compromised accounts~\cite{danezis2009sybilinfer, egele2015towards}, we do not place any additional constraints on the type of spam messages sent nor on the type of accounts used by the adversary.  

Spam detection is an adversarial problem. 
In a real setting, an adversary could reverse engineer how \approach works and actively attempt to evade it. 
Therefore, we assume that the adversary is able to post spam messages through parties of interest similar to those of benign messages. 
Particularly, we assume that the community detection and topic detection algorithms can be played. 
Malicious accounts can compromise some parts of the network, establish connections with honest users and pretend to share the same interests as the target communities. 
We also assume they can replicate the propagation model of benign messages through the parties of interest.
For example, an attacker may observe the number of times a specific benign (viral) message has been posted in compromised communities as well as the number of users who have posted those messages, and then generate or compromise accounts to imitate legitimate parties of interest.

\subsection{Communities and Parties of Interest}

In social networks, users establish connections with others, and through them are able to interact with each other. 
The structure of the underlying connection graph is not homogeneous: rather than being connected with any user in the network, users tend to connect to each other, creating networked communities, where ``everybody knows everybody.''
On the other hand, users tend to come together and form groups to interact around specific topics of interest~\cite{culnan2010large,java2007we,lim2012finding}, \textit{i.e.,} that they tend to form \emph{communities of interest}. 
Thus, the question becomes whether networked communities, defined in terms of the actual connections in the network, coincide with this topic-oriented notion of communities of interest.  

The homophily principle~\cite{mcpherson2001birds} has been observed on online social networks, where users tend to connect to people who hold similar thoughts and values~\cite{weng2010twitterrank, Kwak:2010}.
In summary, this research would seem to suggest that the concepts of community and topics of interest are related.  

However, they cannot be exactly the same. 
Indeed, we expect that in the social network universe there may be several different groups of users that are interested in the same things, but that are not in direct contact with each other. 
In other words, there may exist several networked communities that share the same interest, but that are not connected. 
In that case, the overall community of interest would consist of a set of disconnected networked communities, and would thus not constitute a networked community \emph{per se}. 
Thus, modulo this caveat, one can postulate that networked communities do constitute communities of interest, albeit not a \emph{complete} community of interest regrouping \emph{all} users interested in those topics. 
This is indeed our first hypothesis.

In fact, this hypothesis has been implicitly employed in previous work on communities of interest. 
Indeed, in modern Internet-based social networks it is much easier to determine who is connected to whom than it is to determine \emph{a priori} what are the topics of interest of users, let alone of groups of users. 
Thus, a proxy method for reconstructing these communities of interest has been to extract networked communities from the information on connections between users~\cite{tang2010community}. 
This can be done by using graph algorithms to identify dense subgraphs within the graph of user nodes and connections. 
We explore the validity of this hypothesis on our Twitter dataset. 

\begin{figure}[tbp]
\centering
   \includegraphics[width=1.0\columnwidth]{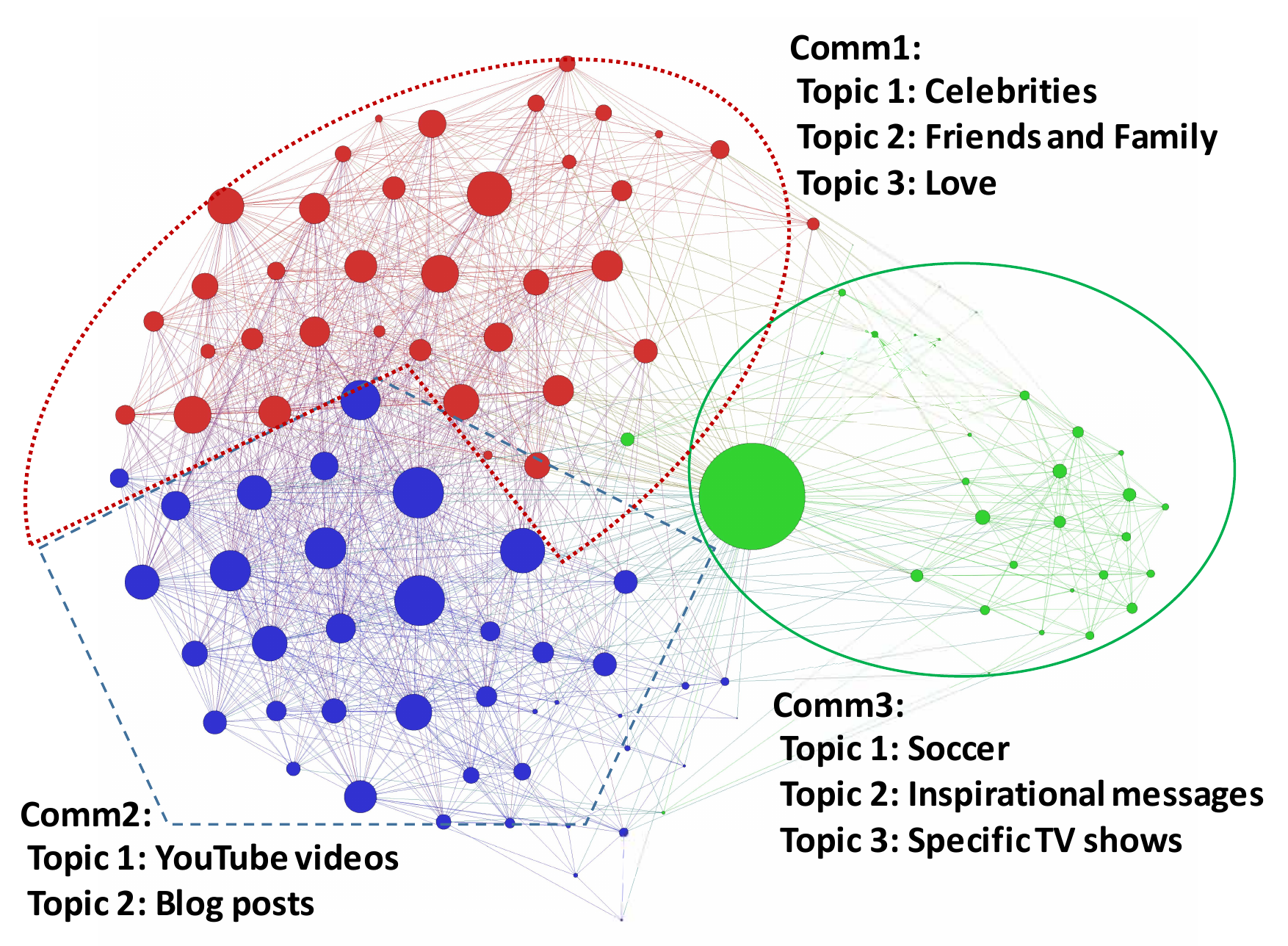}
    \caption{An example of three Twitter communities, where each is talking about a specific set of topics.}
 \label{fig:example}
\end{figure}

In Figure~\ref{fig:example}, we provide an example of three such Twitter networked communities. The edges represent the ``following'' relationships between users. The size of nodes is scaled according to their degree.
We detected these three communities in a subgraph of 2000 users and extracted the topics of their members' discussions. 
We found that each community talks about a specific set of main topics. 
Users in \emph{Comm1} post news about Hollywood celebrities and discuss topics about friends and family. Users in \emph{Comm2} share videos of some specific reputable Youtube users and bloggers. Users in \emph{Comm3} are interested in soccer, post inspirational quotes, and talk about a specific TV show.

Homophily~\cite{mcpherson2001birds} as well as the topology of social networks~\cite{weng2013virality, nematzadeh2014optimal} highly impact the dissemination of information in social networks. 
For example, by analyzing the tweets of communities, it is possible to predict viral memes~\cite{weng2013virality}. 
On social networks, however, not all messages become viral~\cite{lerman2010information} and many just travel through some networked communities interested in similar topics. 
The probability that a certain message is propagated through a set of networked communities is different from another set of communities in the network. 
We call these sets of communities, who care more or less about a message, the \emph{Parties of Interest}. 
We hypothesize that the propagation probability of malicious messages throughout the network distinguishes them from the normal messages.
 
Based on this intuition, we propose a method for detecting malicious content on Twitter, called \approach (Parties Of Interest Semantic Extraction and Discovery). 
\approach computes and learns the propagation probability of messages in the networked communities and extracts the parties of interest.
It then detects malicious messages by distinguishing their propagation probabilities and parties of interest from those of normal messages.
\newpage

%% file: method.tex
\section{Methodology} 
\label{method}

\begin{figure*}[tbp]
\centering
 \includegraphics[width=1.0\textwidth]{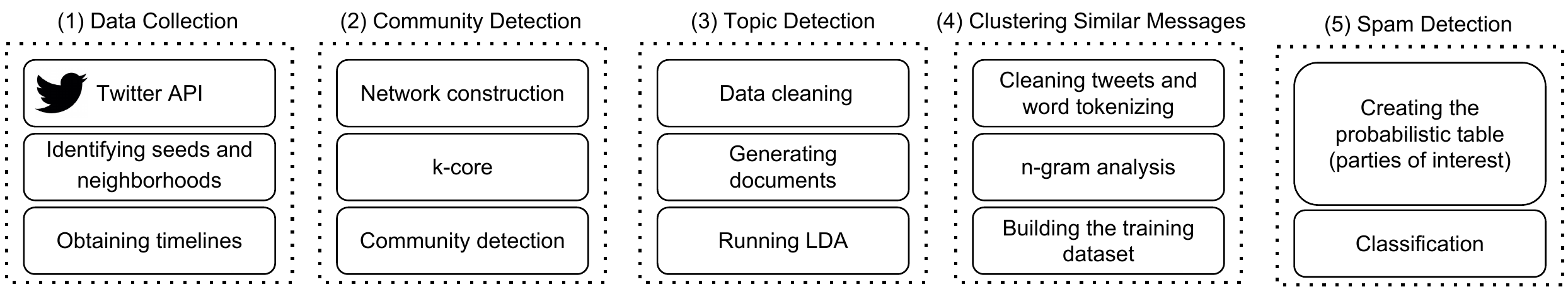}
\caption{\approach constructs a probabilistic model based on the diffusion of messages throughout communities of interest. Then, it employs supervised machine learning to classify messages as spam or benign.}
 \label{fig:framework}
\end{figure*}

The components of our system are shown in Figure~\ref{fig:framework}: 1) data extraction, 2) community detection, 3) topic detection, 4) clustering of similar messages, and 5) spam detection. 
In the following, we explain each component in detail. 

\subsection{Data Extraction}
\label{method-data}
We evaluated \approach on a large-scale dataset extracted from Twitter. 
Twitter is one of the most popular microblogging platforms with over 320 million active users~\cite{twittercomp}. 
This platform enables users to broadcast and share information. 
A user's timeline includes all \emph{tweet} messages posted by that user. 
On Twitter, users follow others or are followed. 
Followers of a user receive all the tweet messages posted by this user. 
Twitter also provides a ``retweet'' mechanism that permits users to spread information of their choice beyond the reach of the original tweet's followers. 
Throughout this paper, we use the terms \emph{messages}, \emph{posts}, and \emph{tweets} interchangeably. 
\approach utilizes user timelines and the social network. 
Here, we formally define a network as:
\begin{definition}
A \emph{social graph} $G \langle V,E \rangle$ is a set of vertices $V$ representing the users in the network and a set of edges $E \subseteq\{(u,v):u,v\in{V}\}$ representing the set of social connections.  
\end{definition}

Note that a social graph $G$ can be directed or undirected. If it is undirected, this means that $(u,v)\in E\iff(v,u)\in E$;  
If, on the contrary, it is directed, then $(u,v)\in E$ does not necessarily mean that $(v,u)\in E$, \textit{i.e.,} $u$ might be connected with $v$ but not vice-versa. 

\subsection{Community Detection}
Although there is no universally agreed-upon definition of a community in a social network, in a graph, structural communities usually refer to a group of nodes that are densely connected to each other and loosely connected to the rest of the graph. 
The nodes inside such a community might also share common properties and/or play similar roles within the graph. 
In social media, communities might have a link with external real entities. 
For example, a user might have a group of friends from the same city, a group from the same school, and yet another group interested in information security. 
Although these communities might be roughly defined and be overlapping with each other~\cite{ahn2010link,Palla05}, the concept behind them is still valid. 
In contrast, however, structural communities are normally defined as disjoint, non-overlapping sets of nodes of the graph.  
In this paper, we will favor the use of structural communities as they are easier to reconstruct from connection information.  
Later, our results suggest that detecting structural communities enables us to detect communities and parties of interest.
Here, we define a networked community as follows: 
\begin{definition}
A networked community structure $C$ of a graph $G$ is a disjoint partition of nodes in $V$, namely $C =\{C_1, \dots, C_h\}$, where $C_i\subseteq V$, $V=C_1\cup \dots \cup C_h$, and $C_i \neq \varnothing$, $C_i\cap C_j=\varnothing$ if $i \neq j$, for all $i, j\in\{1,\dots,h\}$. Nodes in a community $C_i$ are connected to each other with higher probability than to nodes in other communities. 
\end{definition}

\subsection{Topic Detection}
\label{method:topic}
Recently, natural language models have been used for clustering words in order to discover the underlying topics that are combined to form documents in a corpus. 
Topic detection algorithms such as Latent Dirichlet Allocation (LDA)~\cite{blei2003latent} and Topic Mapping~\cite{Lancichinetti2015} have been successfully applied for analyzing text from user messages on social networks. 
By employing a topic detection algorithm, \approach identifies a set of topics of interest for a user and a community. 

The LDA detection algorithm uses a list of documents as an input and detects the corresponding topics. 
For social networks with small message lengths, such as Twitter, topic detection is shown to be less efficient~\cite{hong2010empirical}. For this reason, we aggregate a user's messages into larger documents and then run topic modeling on the documents.
For a user $u$, a set of documents $D_u$, namely $D_u = \{d_1, d_2, .., d_k\}$, is generated by 
partitioning the user's timeline into $k$ groups with $l$ messages. 
Note that $l$ is the number of messages in a document and is a constant, whereas $k$ varies based on the size of the user's timeline. 
Our evaluation with variation of $l=\{1, 5, 10, 20, 50, all\}$ shows that the length of documents do not have a significant impact on the overall results and therefore, we chose $l=20$. 

Formally, a user $u$'s \emph{topics of interest} ($T_u$), namely a list $T = \{t_1, t_2, \dots, t_k\}$, consists of topics detected by a topic detection algorithm on user $u$'s set of documents $D_u$. 

Having extracted the topics for each user's documents, a community's topics of interest can be simply defined as the list of all topics detected for members of that community: 
\begin{definition}
For a community $C$, its set of documents $D_C$ is the union of the documents generated for each community member $u \in {C}$, $D_C = \bigcup _{u\in{C}}{D_u} = \{d_1, d_2, .., d_m\}$, where $m = \sum_{u\in C} |D_u|$.  
\end{definition}

Thus, the set of topics of interest for a community $C$ is defined as follows: 
\begin{definition}
The topics of interest $T_C$ of a community $C$ is a list  $T_C =\{t_1, t_2, \dots, t_m\}$ of $m$ topics detected by a topic detection algorithm on the community's set of documents $D_C$, where each document $d_i$ is labeled with a topic $t_i$.  
\end{definition}

Note that the topics in the topic lists $T_u$ and $T_C$ of a user and a community are not unique, and that several documents for the same user or the same community can be labeled with the same topic. 
Given Hypothesis 1, we expect that the topics found for a user or a community will greatly overlap. 

Detecting topics of interest for each community, \approach constructs a network of structural communities where each community is represented by a set of topics. 

\subsection{Clustering Similar Messages}
\label{method-cluster}
Similarity between messages can be measured by several metrics, and some can be more complex than others~\cite{tsur2013efficient, rosa2011topical, dann2010twitter}. 
Other spam detection methods~\cite{egele2013compa, stringhini2012poultry} have effectively used an approach called \emph{four-gram analysis} to identify similar messages on Twitter. 
This technique proved effective in our case : after manually inspecting the performance of this method on 60 clusters of different sizes, all clusters included the tweets with the same text, \textit{i.e.,} all tweets were correctly grouped. 
As part of this approach, messages that share four or more consecutive words are clustered together. 
While in \approach, other algorithms can also be used to identify similar messages, in our evaluation we cluster messages employing four-gram analysis.

If a message contains less than four words, then all its words in their consecutive order are compared with other messages.
The result of running this algorithm is a list of groups of similar messages, $g= [msg_1, msg_2, ..., msg_j]$, 
where $g$ includes $j$ similar messages, $|g|=j$. The messages in a group can be generated by a single user or several users.

\subsection{Parties of Interest}
A specific message could have been (re-)posted in one or several communities. 
Tracking the propagation of the messages in structural communities identifies \emph{parties of interest}. 
Over time, this tracking makes it possible to predict the probability that a message posted in a community is also posted/retweeted in another community. 
\approach tracks the diffusion of messages over communities, and computes a probabilistic model for every cluster of similar messages.

First, the union set of topics of interest is constructed, which includes all topics detected from all documents. 
Then, for each group of similar messages, \approach counts the number of times that messages in the group have been observed in a community with a specific topic. 
These counts for a group are normalized by the total number of topics identified for messages in that group. 
For example, assume three communities, $C_1$, $C_2$ and $C_3$ are detected in a network, whose topics of interest are $\{t_1,t_2\}$, $\{t_1,t_3\}$ and $\{t_1,t_4,t_5\}$, respectively. 
Assume, a group includes three similar messages posted by users  $u_1$ and $u_2$ in $C_1$, and user $u_3$ in $C_2$. 
The probability distribution for the union of topics in these three communities, $\left(t_1, t_2, t_3, t_4, t_5\right)$, is $\left(\frac{3}{6}, \frac{2}{6}, \frac{1}{6}, 0, 0\right)$.  
All three messages in this group are posted in communities with $t_1$ as their topic of interest ($C_1$ and $C_2$), while only one of these messages is posted in communities with $t_3$ as their topic of interest ($C_2$). Therefore, the count distribution for $\left(t_1, t_2, t_3, t_4, t_5\right)$, is $\left(3, 2, 1, 0, 0\right)$. 
The distributions are normalized by being divided by the message counts of the union of topics, \textit{i.e.,} six in this example, to compute probability distribution.

As a result, for each group of similar messages, a probabilistic model is computed, which shows the potential parties of interest for that group of messages. 
Assume $T$ is the union of all the detected topics in the dataset, where $ T = \{t_1, t_2, ..., t_k\}$. Each group $g$ of similar messages is represented with a topic probability vector $prob_g = [p_1, p_2, ..., p_k]$, where $k = |T|$ and $p_j$ is the likelihood that messages in this group favor communities interested in topic $t_j$;
this probability distribution thus represents the \emph{parties of interest} for messages in $g$. 
The overall table of probability distributions for all groups of messages $prob\_table = \{probs_{g_1}, probs_{g_2},..., probs_{g_h}\}$ thus represents the \emph{parties of interest} for the social network given the observed messages, and it is the basis for the classification model.  
In other words, using this model, if a message is seen in a specific community, it is possible to predict the probability that this (or a similar message) are going to be observed in other communities. 

Note that \approach does not need to learn about the topics of messages, but only about their propagation through communities of interest. 

\subsection{Classification}
If spam messages travel through different parties of interest than those of benign messages, then a classifier can learn these patterns and detect spam messages. 
Hence, by having a ground-truth data set, a classifier can be trained where topics found in communities are features of that classifier, and the class is defined as a binary variable that takes as values: \emph{spam} or \emph{benign}.

%% file: eval.tex
\section{Evaluation Setup}
\label{eval}


\subsection{Dataset}
\label{dataset}

In December 2015, we used the Twitter API to crawl users' timelines. 
The API provides a stream of random users. 
We used a sample of 300 of them, called ``seeds,'' and crawled the timelines for them as well as their friends and followers. 
Thus, we obtained data for 300 \emph{neighborhoods} in Twitter, where a neighborhood consists of a seed with all his friends and followers. 
In our random selection of seeds, we did not collect data for users with more than 2,000 followers or friends, so that the crawling process would be of manageable size. 
We also limited the crawl to users having specified English as their language, so the further text analysis would be performed on tweets of a single language. 
To limit the bias of analysis in favor of older accounts with many tweets, and to work with more current data, we limited the number of tweets used per user to a maximum of 300 of their most recent English tweets.  
In our dataset, the average of all users' oldest and newest tweets are March 2015 and June 2015.

\subsection{Network Construction}
\label{dataset-h1}
For each neighborhood, we constructed a directed and an undirected network based on \emph{following} relationships between all users inside the neighborhood. 
The undirected network is obtained from the directed network, where the relationship between two users must be reciprocal to form an edge. 
In our experiments we examine the impact of both networks on our results.

Some users in each neighborhood only have a single connection to other users. 
These isolated users later result in some communities with a size of one and two. 
We applied a standard technique called $k$-core~\cite{seidman1983network} to extract the maximal connected subgraph of each of the networks, where 
all nodes have a degree of at least $k$, here $k=2$.  

After obtaining the $k$-core of all 300 networks, 
the mean neighborhood size is 271, while some neighborhoods include more than 1000 users. 

\begin{figure}[t]
\centering
\subfigure[Neighborhood size] {\includegraphics[width=0.20\textwidth]{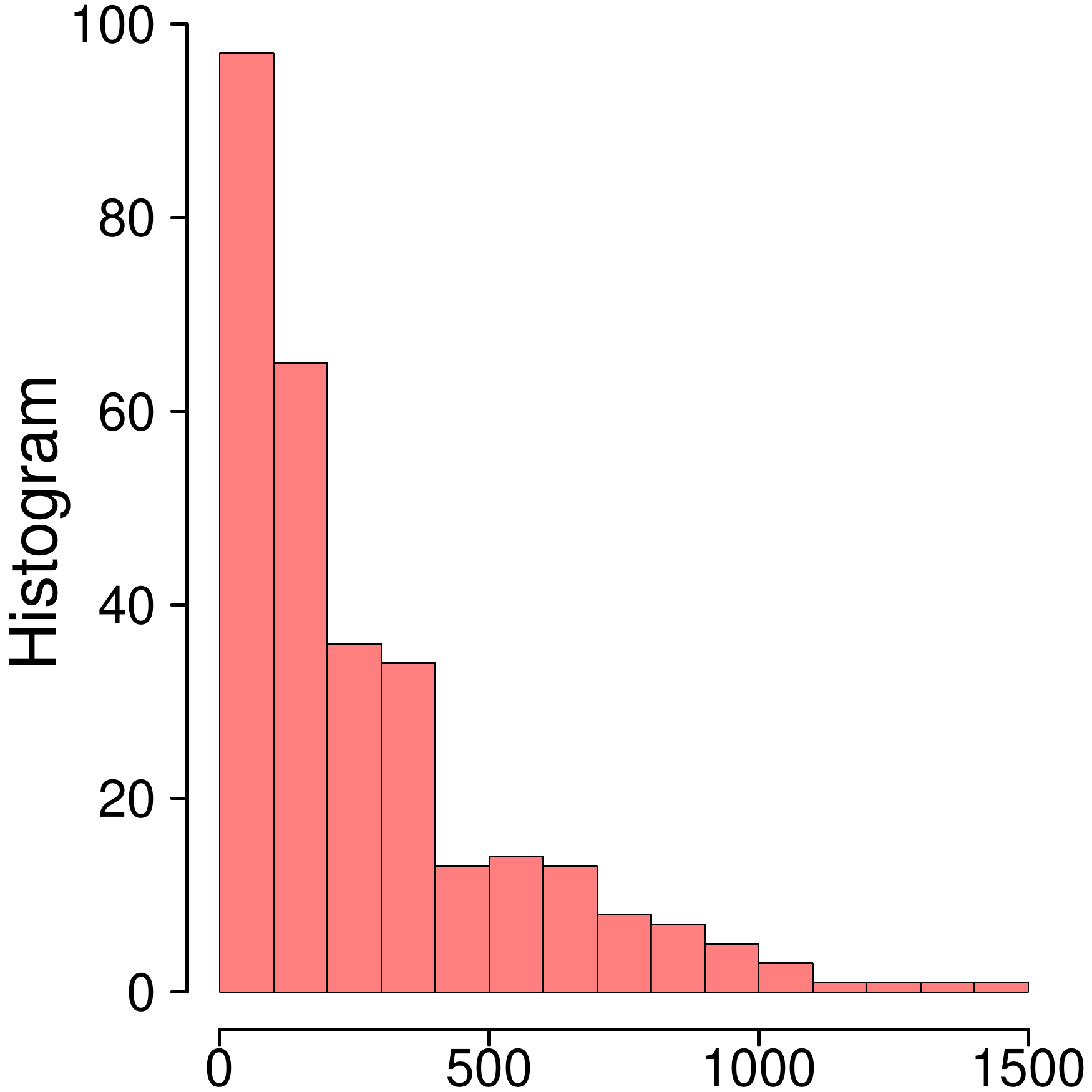}\label{fig:neighborhist}}
\subfigure[Timeline length] {\includegraphics[width=0.20\textwidth]{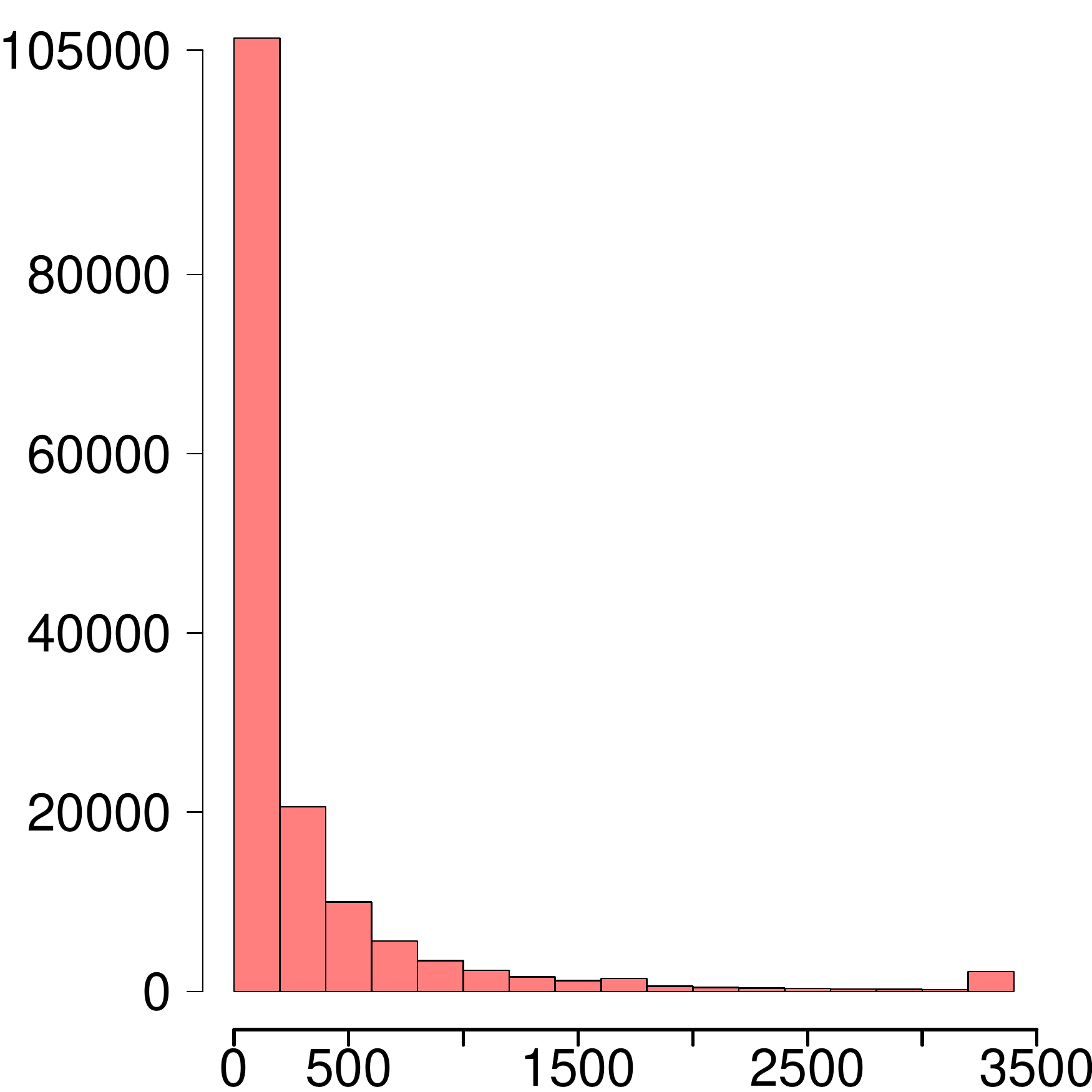}\label{fig:timelines}}
\caption{While on average neighborhoods contain 270 users, some include more than 1000.}
\end{figure}

Figure~\ref{fig:neighborhist} shows the histogram of the number of users (neighborhood size) in 300 neighborhoods.
After obtaining the $k$-core of all 300 networks, the median for neighborhood size is 178 and the mean is 271. 
Seven neighborhoods include more than 1000 users. Figure~\ref{fig:timelines} shows the histogram of timelines' length in our dataset, with an average value of 332 and a median of 177. 
Examining our hypotheses on both small and large neighborhoods demonstrates that our findings are not dependent on the size of a dataset or a specific neighborhood.

\emph{In summary, the dataset for testing the first hypothesis includes 15,751,198 English tweets posted by a total of 82,275 users in 300 neighborhoods.}

\subsection{Community Detection}
To find structural communities, we employed one of the most widely accepted disjoint community detection algorithms, called Infomap~\cite{rosvall2008maps}. This algorithm has shown good performance in tests using benchmark networks~\cite{lancichinetti2009benchmarks, lancichinetti2009community}.\footnote{Among the variety of community detection methods, we evaluated the impact of a set of them on our results, including Infomap~\cite{rosvall2008maps}, Spinglass~\cite{eaton2012spin}, Walktrap~\cite{pons2005computing}, Leading eigenvector~\cite{newman2006modularity}, Fastgreedy~\cite{clauset2004finding} and Multilevel~\cite{blondel2008fast}. 
Interestingly, we obtained very similar results while Infomap slightly outperforms the other algorithms.} 
In Infomap, a community is a partition that minimizes the average number of bits per step required to describe trajectories of random walkers.

\begin{figure}[t]
\centering
  \subfigure[Number of communities] {\includegraphics[width=0.20\textwidth]{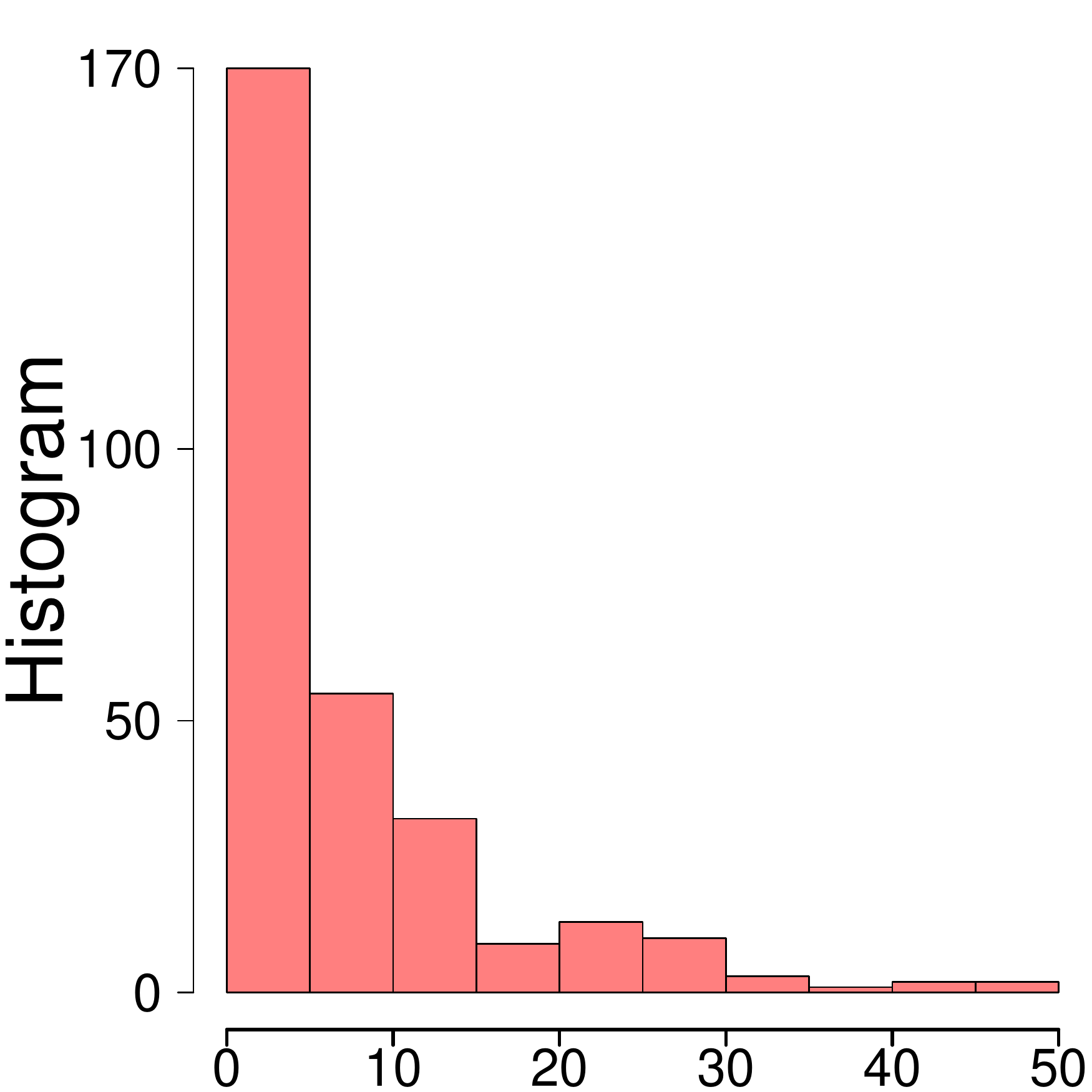}} 
  \subfigure[Size of communities] 
 {\includegraphics[width=0.20\textwidth]{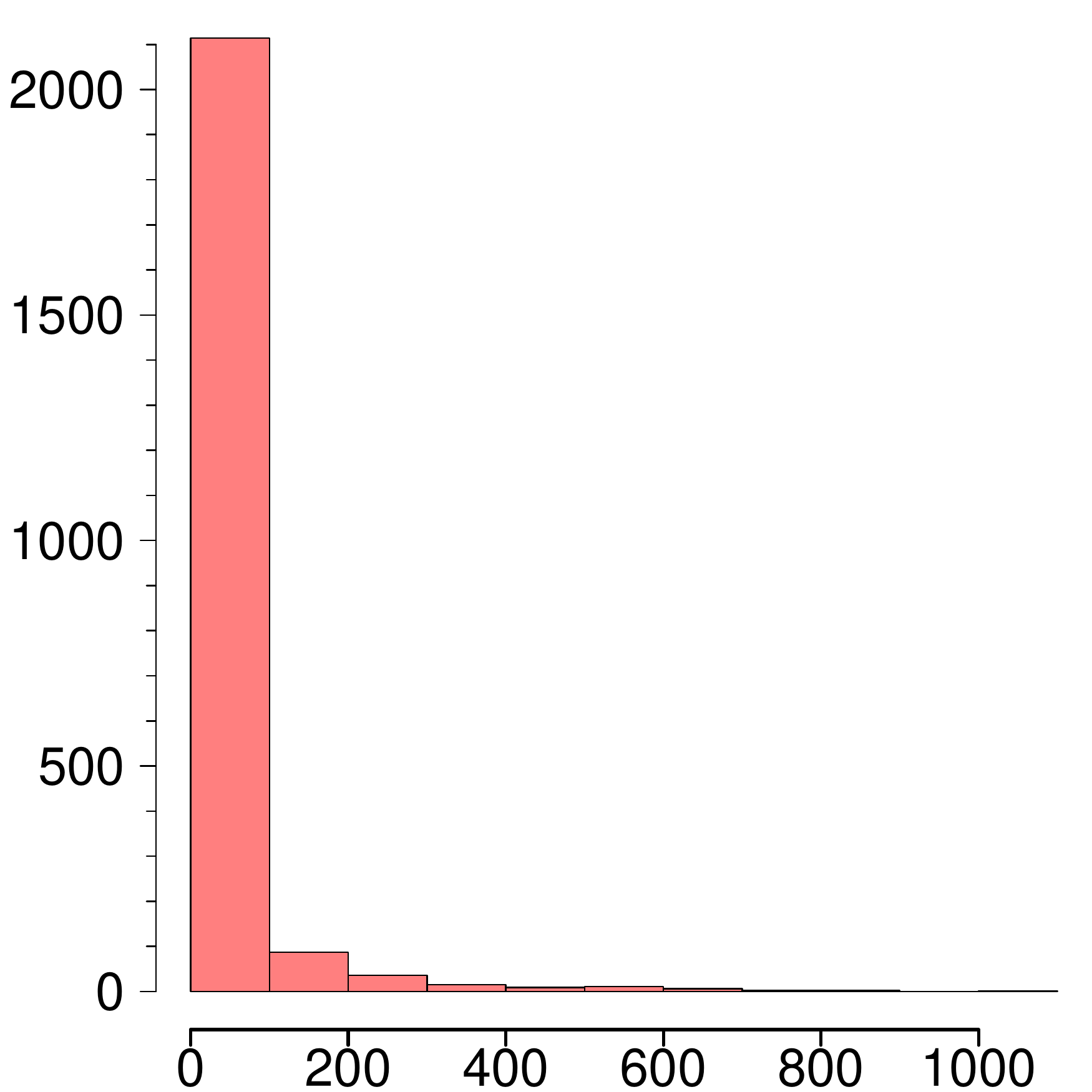}} 
  \caption{Histograms for number of communities and community size.}
 \label{fig:comm:hist}
\end{figure}

Infomap detects a total of 2,283 communities in our 300 neighborhoods. 
While on average neighborhoods contain 8 communities, a handful of them contain more than 30 communities.  
Figure~\ref{fig:comm:hist} shows the histograms for number of communities and their size resulting from the Infomap community detection algorithm. 
The median for number of communities in each neighborhood is 4 and the mean is about 7.5. 
The median for community size is 6 and the mean is about 30. 

 

\subsection{Topic Detection}
We applied Latent Dirichlet allocation (LDA) to find topics of interest for users and communities. 
We first generated text documents from users' timelines and then ran LDA modeling on the documents to obtain a list of detailed topics for each document.

\textbf{Documents}. 
As explained in Section~\ref{method:topic}, 
\approach divides the timeline into several documents.
Since on Twitter, some accounts are older and some users post more often, the lengths of timelines can differ substantially. 
Figure~\ref{fig:timelines} shows the histogram of timelines' length in our dataset, with an average value of 332 and a median of 177. 
On average, the users' timeline include 332 posts. 
As explained earlier, for each user we only consider at most 300 of the most recent tweets for our analysis. 

Each document contains a fixed $l$ number of tweets. 
The number of documents for a user depends on the number of tweets in the user's timeline. 
In our experiments, we investigated the impact of $l$ on the list of detected topics and found that there is not a significant difference with a variation of $l$. Nonetheless, LDA detects topics slightly more accurately with 20 tweets per document. 

\textbf{Topics}. 
We cleaned the documents by removing URLs, and non-printable characters. 
We removed \emph{stop words}, a list of common words found in the English language, to improve topic detection and obtain detailed topics.

We employed the LDA implementation provided by Machine Learning for Language Toolkit (MALLET)~\cite{mccallum2002mallet}. 
The output includes documents labeled with a series of topics. 
We chose to label each document with the topic having the highest weight value.
Having several documents for a user results in possibly several topics for the user. If all the documents of a user are labeled with a particular topic, then it shows that the user is interested in that specific topic. 

MALLET requires a few parameters to apply LDA, such as the amount of topics to be found in the given documents. We experimented with various amounts of topics. Later, we show that the best results are obtained with 500 topics. Moreover, we tested different amounts of tweets per document, and later show that the best results are obtained with 20 tweets per document. In MALLET, we also set the iteration count to 200, which provides more precise topics at the expanse of a longer processing time.

%% file: results-h1.tex
\section{Evaluation: Communities of Interest}
\label{results-h1}
In this section, we examine our first hypothesis that members of a networked community have similar topics of interest. 

\subsection{Metrics and Null Model}
\label{eval_metrics}
We validate Hypothesis H1 on our dataset by computing three entropy-based metrics: \emph{completeness}, \emph{homogeneity}, and \emph{V-measure} of topics detected in communities. 
These metrics were first proposed by Rosenberg and Hirschberg~\cite{rosenberg2007v}, and have been commonly used for evaluating many natural language processing tasks~\cite{rosenberg2007v, csakany1981homogeneity}. 

All these three criteria produce a score in the interval of  $[0,1]$, with 1 being `good' and 0 `bad.'
\emph{Completeness} measures if all documents of a community are assigned to the same topic, \textit{e.g.,} they are only about football. 
Symmetrically, \emph{homogeneity} measures if each topic is only observed in a single community, \textit{e.g.,} if all messages about a local art competition are posted by members of one community. 
\emph{V-measure} is the harmonic mean of completeness and homogeneity and measures how successfully the two criteria are met. The computation of these measures is independent of the number of documents and topics in the communities and the network.


By computing these three metrics, we are able to measure if the members of a detected community are interested in the same topic. 
%

To further validate that communities provide additional information about members' common interests, we propose comparing their scores with those of a \emph{null model}. 
Statisticians use \emph{null models} as baseline points of comparison for assessing goodness of fit
~\cite{perry2012null}. Recently, null models have been used for studying network structures~\cite{perry2012null, newman2003social, Cho:2011:FMU, su2013null} and community structures~\cite{mucha2010community}. 
We generate a \emph{null model} that randomly partitions documents into groups. 

To have a fair comparison between the documents in the actual communities and those of the random clusters of the null model, the documents are shuffled so that the distribution of size of groups in the null model remains the same as in the detected communities in the network, and the number of communities remains the same. 
As a result, if the actual communities have higher scores, then our hypothesis is validated that users in communities are grouped over some specific topics of interest, and random clustering of documents does not provide similar or better scores.

\subsection{Communities Discuss Different Matters While Community Members Talk About Similar Topics}
We examine Hypothesis H1 by comparing the communities of interest that are detected in our neighborhoods with the random communities of the null model. 

\begin{figure}[tbp]
\centering
 \includegraphics[width=0.40\textwidth]{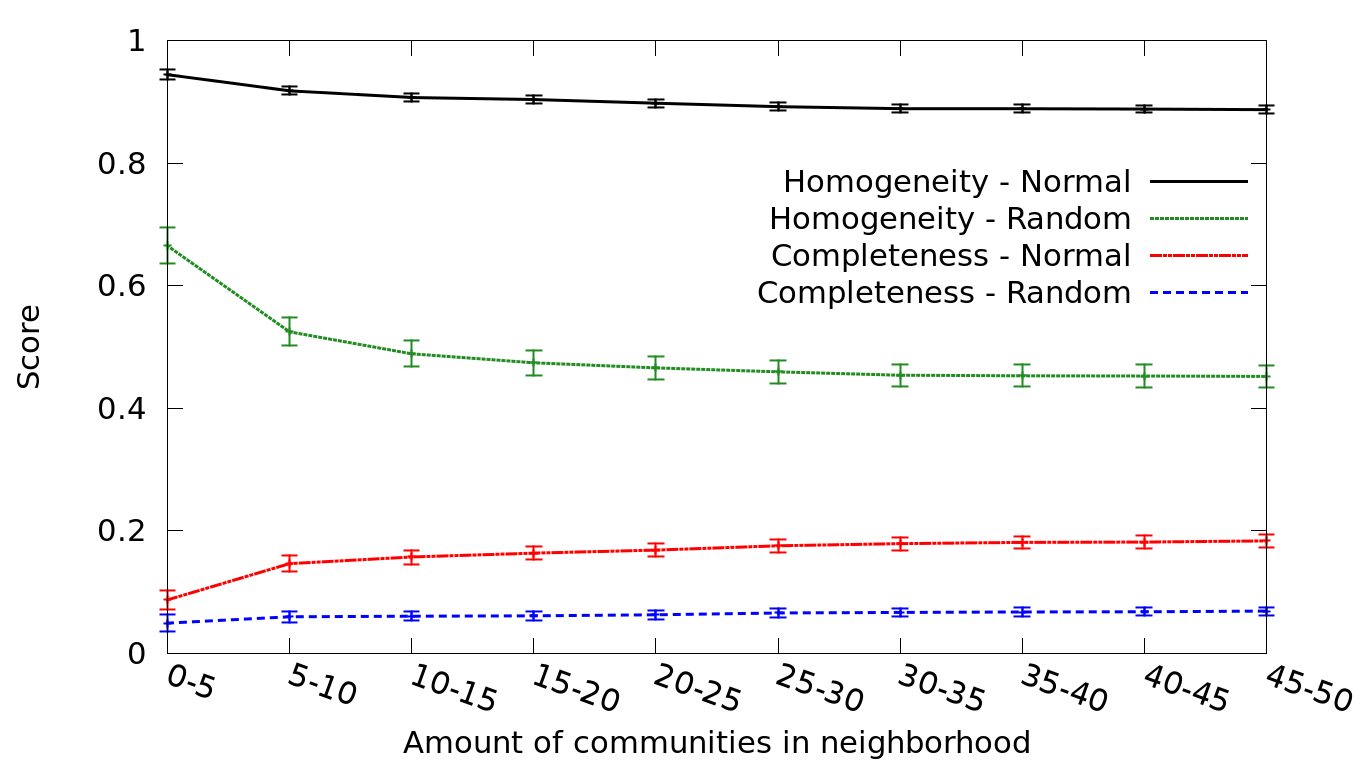}
 \caption{Both metrics highly decrease for the null model.}
 \label{fig:rand_comms}
\end{figure}

Figure~\ref{fig:rand_comms} compares the scores for these two models. To examine the effect of the number of communities on the scores, Figure~\ref{fig:rand_comms} presents them according to the number of communities in the neighborhoods.  
The \emph{normal} results denote the scores obtained from the model created from actual communities, while the \emph{random} results show the results for the null model created from random communities. 

Overall, the tests with random communities show a substantial drop in results. For communities of interest, on average, completeness and homogeneity scores are about 0.16 and 0.90 respectively while those scores are about 0.063 and 0.49 for random communities. 
We also ran a \textit{Z-test} to compare the homogeneity and the completeness values for actual and random communities. For both metrics, the p-values are lower than 0.0001.    

\emph{The completeness and homogeneity scores indicate that people in the communities talk about certain topics that are mainly different from those talked in other communities in their neighborhood. The homogeneity results also confirm that communities of interest are effectively distinguishable.}
 
\begin{figure}[t]
\centering
 \includegraphics[width=0.39\textwidth]{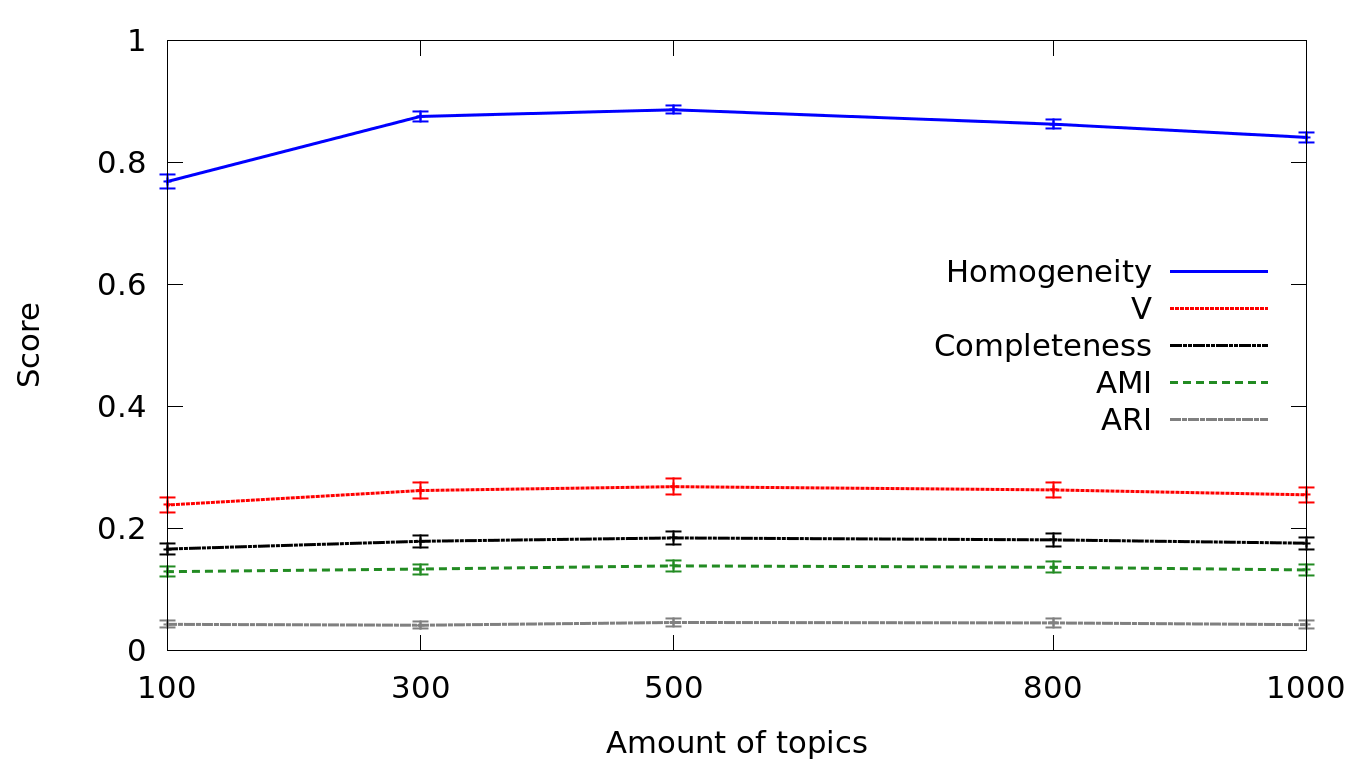}
  \caption{The metrics' score slightly increase for 500 topics.}
 \label{fig:amount_topics} 
\end{figure}

Some parameters can affect our evaluation of the first hypothesis. 
For example, the total amount of possible topics is a fixed value given as a parameter to MALLET. Also, we used a fixed amount of tweets written into each document. 
Figure~\ref{fig:amount_topics} shows the impact of the amount of topics varying from 100 to 1000.  
As can be seen, the amount of topics do not significantly change the scores. 

Only the homogeneity score slightly varied, being 0.87 at its maximum, when the topic count is 500.
Similarly, we tested the impact of document size varying from 5 to 300 tweets per document. In summary, this parameter only affects the scores slightly (at most 0.17 in the range of [0,1]) and the best scores are provided when the document contains 20 tweets. 

To measure the impact of different community detection algorithms, we also re-run the experiments for communities detected by several of these algorithms. 
The topic detection by LDA is independent of the community detection algorithm, and each community includes topics detected for its members. 

\begin{figure}[t]
\centering
 \includegraphics[width=0.47\textwidth]{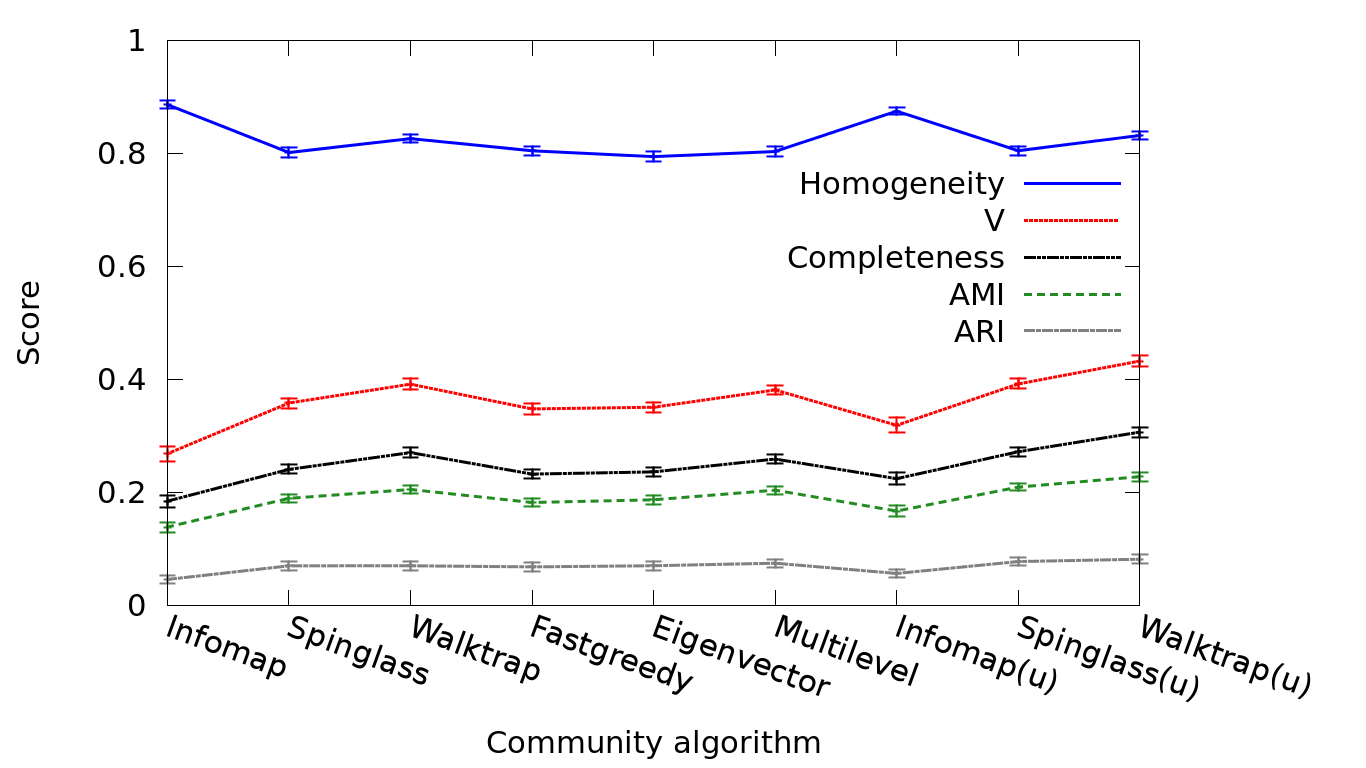}
  \caption{The communities and their topics of interest show high homogeneity but relatively low completeness.} 
 \label{fig:comms}
\end{figure}

Figure~\ref{fig:comms} shows the scores for various community detection algorithms. These scores were calculated for each neighborhood, and then averaged. 
It illustrates that no matter what community detection algorithm is used, the communities and their topics of interest demonstrate high homogeneity, $[0.8, 0.89]$. 
Higher homogeneity confirms that communities do have little topics in common and, hence, are distinguishable. Considering its high homogeneity score, \emph{Infomap} was chosen to be applied by \approach. 
Similarly, no matter what community detection algorithm is used, the completeness between communities and their topics is about 0.25.  The \emph{Walktrap} community detection algorithm provides communities with the highest completeness, with a score of 0.31. These values show that, for communities, not one but multiple topics are detectable. 

Since the scores for the directed networks are slightly higher, we used the directed networks for our analysis.

%% file: results-h2.tex
\section{Evaluation: Twitter Spam Detection}
\label{results-h2}
Here, we examine our second hypothesis that benign and malicious content diffuse through distinguishable parties of interest, which can be used to detect Twitter spam messages.   

\subsection{Clustering Similar Messages}
\label{similar_msgs}
First, \approach needs to observe the diffusion of messages through communities of interest to learn about their parties of interest. 
For this, it applies four-gram analysis to detect similar messages in every neighborhood. 
We cleaned tweets by removing stop words and punctuation. 
Also, each URL is considered as a word. 

\begin{figure}[t]
\centering
\subfigure[Global dataset] {\includegraphics[width=0.21\textwidth]{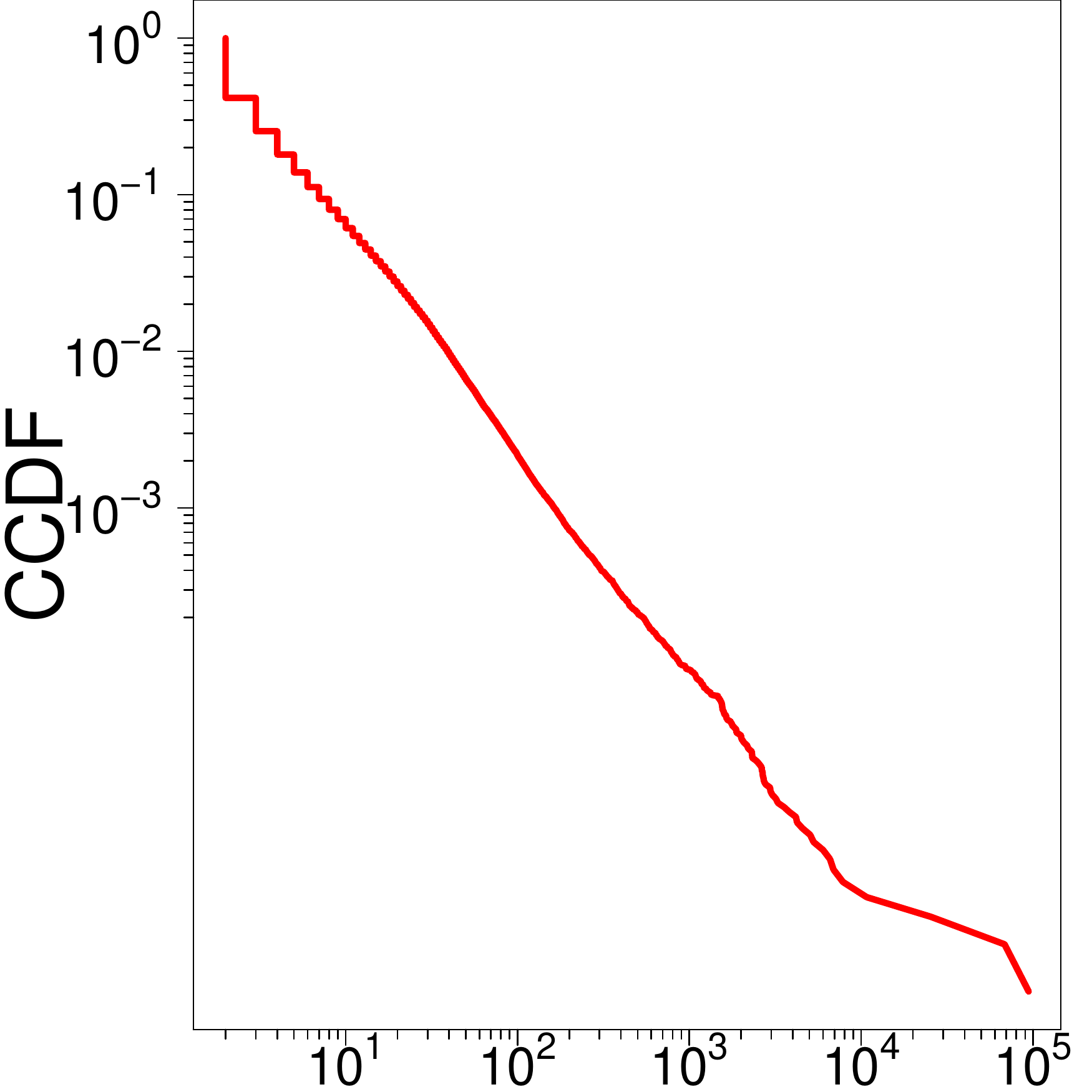}\label{fig:group_size}}
\subfigure[Labeled dataset] {\includegraphics[width=0.21\textwidth]{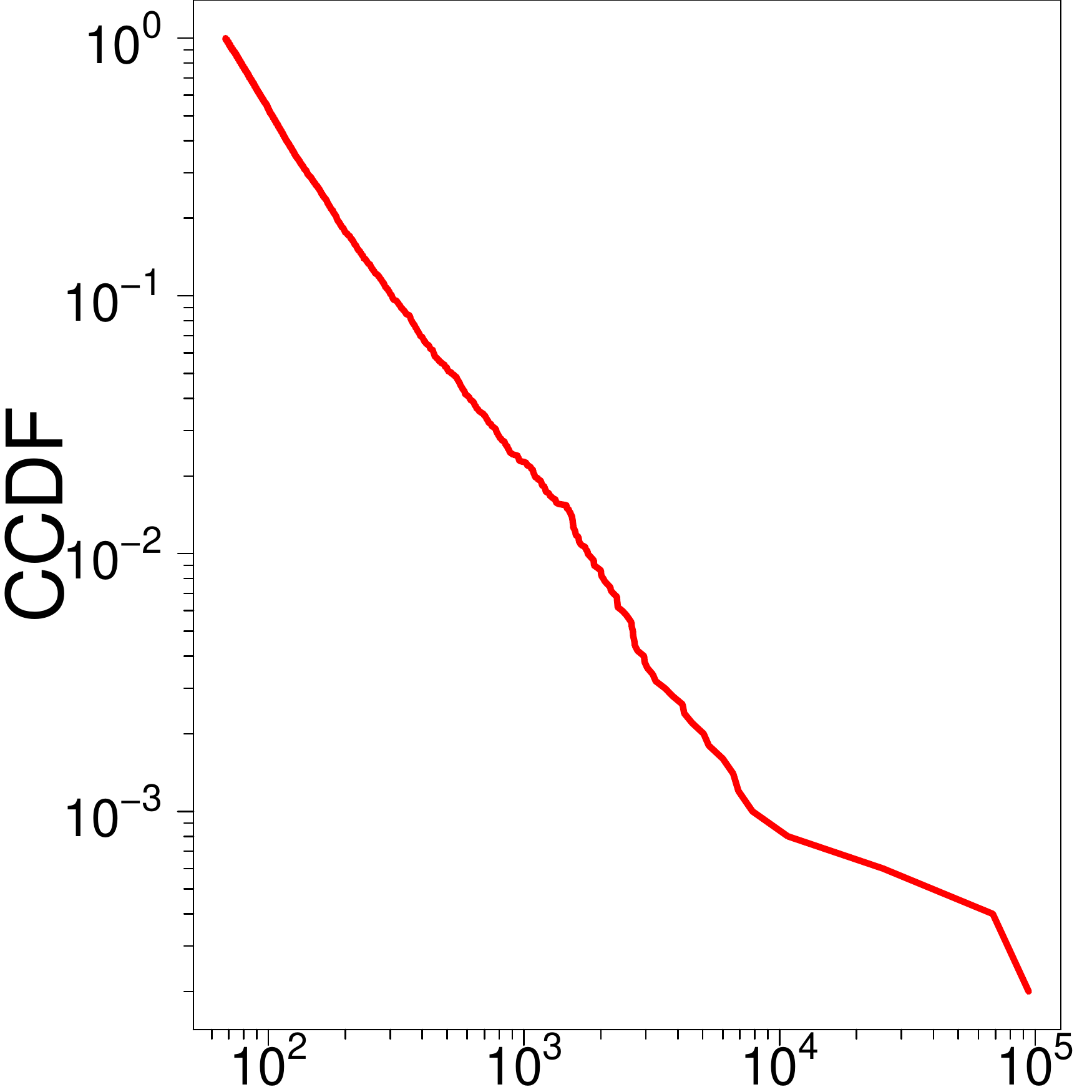}\label{fig:group_size_ground}}
\caption{Size of groups of similar messages in the whole dataset and labeled dataset.}
\end{figure}

In total, in our dataset, we found 1,219,991 groups of similar messages with the size range being between 2 and 94,382. 
Figure~\ref{fig:group_size} shows the CCDF of the size of groups of similar messages in our dataset, which follows a power law distribution, with a small number of big groups and many small ones.

\subsection{Create a Labeled Dataset}
\label{eval-ground-truth}
To evaluate the probabilistic models for spam detection, we need to have a ground-truth dataset including both spam and benign messages. 
Since it is not possible to manually label over 1 million clusters of similar messages, we picked the top 5000 groups after ordering them by their size, and labeled these groups. 
The group size picked for labeling ranges from 68 to 94,382.
We may find more malicious campaigns by looking at larger clusters compared to all data. 
However, many large clusters also contain benign messages such as simple ``happy birthday'' wishes, as well as memes and trending topics. 

The tweets in this dataset were manually checked by a group of 14 security researchers who labeled them independently, following a similar methodology to the one applied by previous work~\cite{chatzakou2017mean}. 
The researchers were advised on the risks of clicking on suspicious links and took precautions not to get infected (\textit{e.g.,} by using virtual machines to lookup URLs). 
Each group of similar tweets is evaluated and labeled by three researchers, and then the majority vote is considered as the final label. 
We provided some guidelines for the researchers and defined some categories so that they can label each tweet with one of those categories. These categories are defined as: 

\begin{table*}[t]
\centering
{\small  
  \caption{Examples of messages in each category} 
  \label{example-categories}
  {\renewcommand{\arraystretch}{2}%
\begin{tabular}{ll}
\hline
\hline
\textbf{Spam} &  \begin{minipage}{0.70\textwidth}``Fellas need a mix 2 get ur lady in the mood heres a mix to help u succeed. [\_URL]''\\
``@X Listen To My Lul Song XXX Hot! Download And Share [\_URL]'' \\ 
 ``Wow! another great item; available on eBay [\_URL]''\\
 ``WOW! No Cost Traffic For Your Website Home Based Business Blog Click Here Now Please \#retweet [\_URL]'' \end{minipage}\\
\hline
\textbf{App-generated} & \begin{minipage}{0.70\textwidth}``4 tweeps unfollowed (goodbye!) me in the past week. Thank you [\_URL].'' \\ ``New week; new tweets;  new stats. 2 followers; 3 unfollowers. Via good old [\_URL].''\\
``August 28; 201X \#Fitbit activity: 11XXX steps taken; 5.XX miles walked/ran; and 2XXX calories burned.''\\
 ``Your key planet Venus is now moving through your 12th House of... More for Libra  [\_URL]''\end{minipage} \\
\hline
\textbf{Quote} & \begin{minipage}{0.70\textwidth}``Either you run the day or the day runs you. - Jim Rohn'' \\ 
``You're only as good as the people you hire. - Ray Kroc''\\
``Do you want to know who you are? Don't ask. Act! Action will delineate and define you. - Thomas Jefferson''\\
``Problems are only opportunities in work clothes. - Henry J. Kaiser'' \end{minipage} \\
\hline
\textbf{Normal} & \begin{minipage}{0.70\textwidth}``If you could ask a business consultant any question;  what would you ask?''\\ ``Brendan Rodgers: Liverpool boss has no plans to leave club [\_URL]''\\
``RT @X: Thank you XX for last night. Hope you all enjoyed the show;  you've always been lovely to us.''\\
``RT @X: Puppy caught eating paper decides killing the witness is the only way out [\_URL] \end{minipage}\\
\hline
\end{tabular}}}
\end{table*}

\noindent\textbf{Spam}: A message that is encouraging users to do something, such as buying an item, voting for someone, visiting a URL, etc. 
Encouraging users is not by itself a malicious activity, however, if a tweet is doing it, then a more careful assessment is needed. 
For example, coders had to visit the websites these URLs refer to. 
If the URL is suspicious, then the message is spam. 
Note that memes also may include URLs. 
If in some cases, the URLs are not functional, then the coders were advised to label the groups based on their subject, and their tone.

\noindent\textbf{App-generated}: A message that is automatically posted by an application on the user timeline. 
Some examples are weather alerts posted by IFTTT\footnote{https://ifttt.com} and health-related reports posted by fitness trackers such as Fitbit.\footnote{https://www.fitbit.com} In the process, we also found that some of the bigger clusters are the result of some apps such as Twittascope, which regularly post tweets on behalf of the users. These tweets usually contain links to some articles.  

\noindent\textbf{Quote}: A message that is a popular quote. In the first round of labeling, we found that many tweets consisted of quotes, so we created a separate category for them, and asked the researchers to fix their labels accordingly.

\noindent\textbf{Normal}: A message that seems normal and has become popular (trending) because of its content. 
Examples are memes about current news and links to interesting reads, videos, photos, etc.

\noindent\textbf{Unknown}: If the coders were unsure about the category of a message, they could choose 'Unknown.' However, researchers were advised to try not to choose this option.

Table~\ref{example-categories} provides some examples of manually labeled messages. 
Table~\ref{manual-labeling} shows the size of each category after labeling all 5,000 clusters of similar messages. 
It also shows the number of tweets in each category. 
In total, by labeling these 5,000 groups, we obtained labels for 1,277,833 tweets. 
As you can see, clusters of similar messages are almost evenly labeled as normal (44\%) and spam (42\%). Most tweets though are labeled as normal (38\%) because groups labeled as normal include more tweets. 
While only about 8\% of groups are labeled as app, they include about 33\% of tweets. 

Labeling the tweets also can be utilized to classify users into two groups of spam and benign users. 
Table~\ref{manual-labeling} shows the amount of unique users identified in each category. 
The total amount of unique users in all the groups is 66,788. 
However, some users appear in multiple categories. 
For users who appear in both spam and normal categories, three possible explanations are: 
First, spam accounts may try to emulate normal users to avoid suspension by Twitter; second, accounts may have been compromised, and, therefore, some posts on their timeline are benign while others are spam; and, third, their tweets are mislabeled. 

Another interesting observation is that the amount of tweets in the normal category is only about 5\% and 10\% more than that in the spam and app categories, respectively, while the number of unique users in the normal category is almost 4 and 6 times more than that in the spam and app categories. 
These considerable differences indicate that spam accounts or campaigns are responsible for larger clusters that repeatedly post similar spam messages.

\begin{table*}[tbp]
\centering
{\small  
  \caption{Statistics for the manually labeled and ground-truth datasets} 
  \label{manual-labeling}
\begin{tabular}{lccccc}
\hline
\hline
& \textbf{Spam} &  \textbf{App} & \textbf{Quote} & \textbf{Normal} & \textbf{Unknown} \\
\hline
\multicolumn{6}{c}{Labeled Dataset (300 neighborhoods)}  \\
\hline
No. of groups & 2,110 (42.2\%) & 376 (7.5\%) & 335 (6.7\%) & 2,178 (43.6\%) & 1 (0\%) \\
No. of tweets & 344,540 (27\%) & 416,099 (32.5\%) & 34,138 (2.7\%) & 482,975 (37.8\%) & 81 (0\%) \\
No. of users & 13,179 & 9,740 & 3,819 & 55,473  &  1 \\
\hline
\multicolumn{6}{c}{Ground-truth Dataset (202 neighborhoods)} \\
\hline
No. of groups &  854 (30\%) & 274 (9\%) & 260 (9\%) & 1508 (52\%)\\
No. of tweets & 168,181 (17\%) & 408,395 (40\%) & 32,607 (3\%) & 408,340 (40\%) \\
No. of users & 12,504 & 9,614 & 3,815 & 55,219 \\
\hline
\end{tabular}}
\end{table*}

\subsection{Identifying Parties of Interest}
\label{evl-prob-table}
The probabilistic table representing the parties of interest is generated using the labeled dataset. 
As it was explained in Section~\ref{model}, for each cluster of similar messages, \approach computes the probabilities of messages in that group being posted in each of the communities of interest. 
As an example, let us assume that in a neighborhood with three communities, the topic detection algorithm has found eight topics of interest and four-gram analysis has identified ten clusters of similar messages. 
The probabilistic table will then include ten rows and eight columns corresponding to groups and topics, respectively. 
The table entry for row $i$ and column $j$ is the probability that the messages in group $i$ is being observed in communities with an interest in topic $j$. 

Ideally, the probabilistic model should be built on the whole data of a social network. 
However, we cannot perform topic modeling on all data because of the resource constraints imposed by the use of MALLET. 
Instead, the topic detection is run and a probabilistic table is generated for each neighborhood separately. 
Each neighborhood includes multiple communities. 
We show that even while performing the analysis locally on each neighborhood, \approach detects spam messages effectively. 
This technique of running the analysis on neighborhoods can also help scaling our approach, so that it can be applied on much larger datasets. 
For example, Twitter can divide its large dataset into a few partitions that are then independently analyzed. 
 
A probabilistic table for a neighborhood includes the clusters of similar messages that are posted in that neighborhood. 
Also, if messages of one large group are observed in multiple neighborhoods, then the probabilities for this group is computed separately and listed in the probabilistic table of every neighborhood. 
As a result, the size of probabilistic tables varies for each neighborhood. 
Some include thousands of clusters of similar messages while others include only a few of them. 
Because of the lack of enough observations in some neighborhoods, we did not run \approach on neighborhoods with less than 10 benign or 10 spam clusters of similar messages. 

\emph{Therefore, our `ground-truth' dataset for testing the second hypothesis includes the data for 202 neighborhoods with 2,896 clusters of similar messages and close to 1.3M tweets generated by more than 64K users.}

Table~\ref{manual-labeling} summarizes some statistics on this dataset. 
It shows that the number of spam groups and tweets in the ground-truth dataset are about 12\% and 10\% less than those in the `labeled' dataset. 
In contrast, the number of app groups and tweets in the `ground-truth' dataset are about 2\% and 8\% more than those in the `labeled' dataset.

\subsection{Classification on Parties of Interest}
\label{classification}
We employed machine learning to detect spam messages. 
The features are the list of topics in a neighborhood that are automatically detected by the topic modeling algorithm. 
An observation indicates the parties of interest represented by a row in the probabilistic table that includes the probabilities that a group of similar messages has been observed in communities interested in each of the topics. 
In addition to these features, we also consider the number of users per total number of messages in a group as a feature. This feature aims at capturing if a message is posted by several users or only by a small number of them. 

The class of a group is the label in the ground-truth dataset. 
We define a class as a binary variable: `benign' or `spam'. 
While five categories were defined for manual labeling, for evaluation, we examined different combinations of these categories: 
\emph{Comb. 1}: $Spam = \{spam\}$, and $Benign = \{normal, quote, app\}$, \emph{Comb. 2}: $Spam = \{spam\}$, and $Benign = \{normal\}$ and \emph{Comb. 3}: $Spam = \{spam, app\}$, and $Benign = \{normal, quote\}$. 
Later, we show that regardless of the combination, `spam' messages are detected with high accuracy. 
The distinction between spam and benign messages can be specified by a policy and fed to \approach as a parameter.

The datasets are not balanced, \textit{i.e.,} the number of spam and benign messages is not equal. 
We overcome this limitation by using a well-known over-sampling technique called \textsc{SMOTE} in which the minority class is over-sampled by creating ``synthetic'' examples~\cite{chawla2002smote}. 

To assess the effectiveness of our spam detection algorithm, we use the standard information retrieval metrics including recall, precision, F1-score, and accuracy. 

After creating all the probabilistic models over the ground-truth dataset, we applied $k$-fold cross validation on each of the 202 neighborhoods separately. 
Then, we averaged their measures. We tested with three values of $k= \{3, 5, 10\}$ and found very similar results. Since $k=10$ is one of the most common practices~\cite{kohavi1995study, refaeilzadeh2009cross, braga2004cross}, we report the results for that value. 
We also tested several classification algorithms including Naive Bayes, SVM and Random Forest and all provide similar results. 

Table~\ref{detection-results} shows the results provided by SVM on different observation combinations. 
Since most groups of messages are labeled as `spam' or `normal', these combinations do not highly impact the results of the classifier. 
The results suggest that with high precision (91\%) and recall (93\%) our classifier successfully detects spam messages.

\begin{table*}[tbp]
\centering
{\small  
    \caption{The performance of \approach as well as state of the art systems. The +/- values indicate standard error.}  
  \label{detection-results}
\begin{tabular}{lc|cccc}
\hline
\hline
\textbf{System} & \textbf{Label Combination} & \textbf{Accuracy} & \textbf{F1-score} & \textbf{Precision} & \textbf{Recall} \\
\hline
\approach  & Comb. 1 & 0.89 (+/- 0.02) & 0.90 (+/- 0.02) & 0.85 (+/- 0.02) & 0.95 (+/- 0.02) \\
\approach & Comb. 2 & 0.90 (+/- 0.02) &  0.91 (+/- 0.02) & 0.87 (+/- 0.02) & 0.95 (+/- 0.02) \\
\approach & Comb. 3 & 0.90 (+/- 0.02) &  0.90 (+/- 0.02) & 0.91 (+/- 0.02) & 0.93 (+/- 0.02) \\
\hline\hline
\textsc{SpamDetector}~\cite{stringhini2010detecting}
& Comb. 3 & 0.67 & 0.20 & 0.21 & 0.20 \\
\textsc{COMPA}~\cite{egele2013compa} & Comb. 3 & NA & 0.55 & 1.0 & 0.38 \\
\textsc{BotOrNot}~\cite{davis2016botornot} & Comb. 3, thr=0.8 & 0.76 & 0.07 & 0.36 & 0.04 \\
\hline
\end{tabular}}
\end{table*}

%

\subsection{Spam Accounts in Labeled Dataset}

\begin{figure}[tbp]
\centering
 \includegraphics[width=0.25\textwidth]{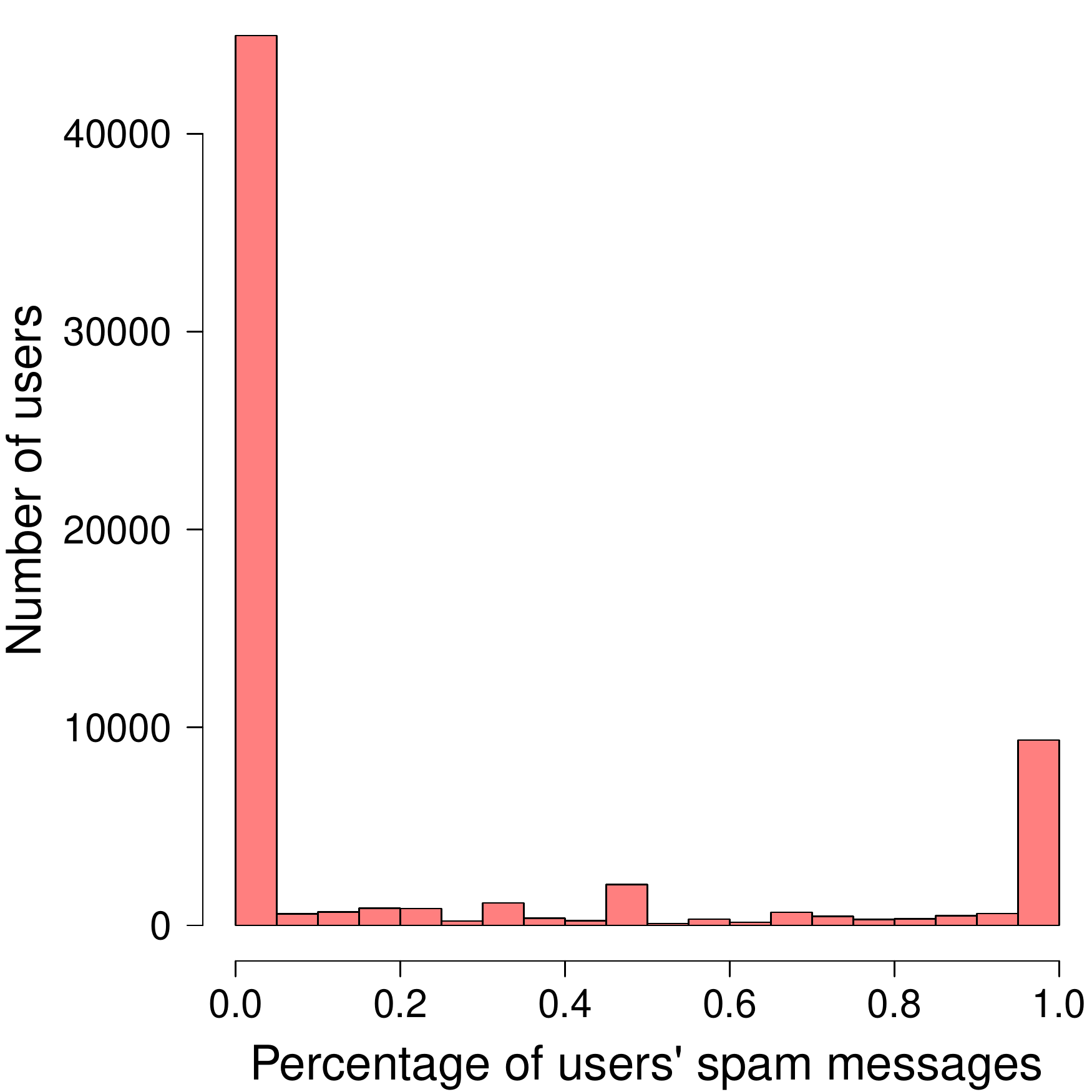}
 \caption{Percentage of spam posted by individual users.}
 \label{fig:per-spam}
\end{figure}

While some users only posted either benign or spam messages, some users have posted messages from both classes. 
Figure~\ref{fig:per-spam} indicates the histogram of spam message percentages for all users in the labeled dataset. It illustrates two main clusters of users who either post benign or spam content. This confirms that by examining the distribution of an account's messages, it is possible to label an account as benign or spam. 

To identify spam accounts, we compute $s$, the percentage of spam messages to the total number of messages that a user has posted. 
%
%
A user is identified as a spam account if $s \geq \tau$. $\tau$ is a parameter that can be configured based on the dataset and the desired or accepted false positive values for the system. 
We set $\tau=0.4$ and assume that if more than 40 percentage of the messages posted by an account are spam messages, then this account is more likely a spam account.

Finally, our labeled dataset includes 15,055 (23\%) spam accounts and 49,675 (77\%) honest users. The high percentage of spam accounts may be due to our selection of larger groups of similar messages as a labeled dataset, which may relatively include more spam messages compared to the whole Twitter data. 

\subsection{Comparison with state of the art systems}

We compare \approach to three state-of-the-art systems that have been proposed by the research community in the past:  
\textsc{SpamDetector}~\cite{stringhini2010detecting}, which is a system that detects fake accounts on social networks by examining characteristics of these
profiles (\textit{e.g.,} the fraction of messages posted that contain a URL), 
\textsc{COMPA}~\cite{egele2013compa}, which detects social network accounts that have been compromised by learning the typical behavior of an account and
flagging any activity that deviates from that behavior as a possible compromise, and \textsc{BotOrNot}~\cite{davis2016botornot,ferrara2014rise}, which leverages more than one thousand features to evaluate if a Twitter account exhibits similarity to the known characteristics of social bots. 
We could not compare \approach with a couple of more recent work due to either the difficulty in obtaining their systems, or not being applicable on Twitter data. We discuss them in more details in Section~\ref{related}.

Note that the threat model tackled by our approach is much broader than the one that these systems focused on:
we aim to detect any malicious message regardless whether it was posted by a
fake account or by a compromised one, while previous approaches only focused on
one of these two categories. Because of this reason, our results in Table~\ref{detection-results} show that our system outperforms both \textsc{SpamDetector} and \textsc{COMPA}, as well as \textsc{BotOrNot}.

\textsc{SpamDetector} and \textsc{COMPA} were developed as part of previous work by some of the authors of this paper, therefore we had access to their source code. 
In the case of \textsc{SpamDetector}, we performed a 10-fold cross validation on our labeled dataset using Random Forest as a classification algorithm. 
For \textsc{COMPA}, performing a 10-fold cross validation would not make sense,
since this system does not take into account two classes of spam or benign accounts, but rather learns the typical behavior of an account and determines whether new messages that an account sends are malicious or not. 
Therefore, in this experiment we used \textsc{COMPA} to learn the typical behavior of the accounts that sent spam in our ground-truth dataset, and determine whether the spam messages that they sent were the consequence of a compromise. 

Table~\ref{detection-results} illustrates that our system outperforms both \textsc{COMPA} and  \textsc{SpamDetector} considerably. 
The perfect precision reported by \textsc{COMPA} is an artifact of the fact that we only tested that system on malicious accounts (which is also the reason why we could not calculate accuracy), but the low recall of 0.38 (possibly due to the fact that only a minority of the accounts in our labeled dataset were compromised and not just fake) shows that \approach is a better candidate to fight the problem at hand.

We called the \textsc{BotOrNot} API for all the users in our labeled dataset. The classification system of \textsc{BotOrNot} is based on more than 1,000 features extracted from interaction patterns and content. 
When testing on a user, it returns the probability of that user being a social bot (Sybil account). 
To label users as bots, we picked multiple values, $\{0.7, 0.8, 0.9\}$, as a threshold. Here, we reported the values of measures for 0.8. 
We ran \textsc{BotOrNot} in February of 2016. 
Since both tools were run relatively close to each other, 60550 out of 63600 accounts still existed and were accessible for \textsc{BotOrNot}.  
In all cases, \approach outperforms \textsc{BotOrNot}. 
Overall, the precision of \textsc{BotOrNot} is around 40\% while the recall is as low as 0.03. 

\subsection{Missed Spam Messages and Accounts}
All of the above systems are relying on classifiers that find the abnormalities based on some features that can be adopted over time. 
The systems that we compared against \approach are either based on specific characteristics of the accounts under scrutiny (\textsc{SpamDetector}, \textsc{BotOrNot}) or look for changes in the behavior of an account that might be indicative of a compromise (\textsc{COMPA}). 
Given these peculiarities, these systems can only detect certain types of spam. 
In contrast, the feature set in \approach is adopted over time by detecting parties of interest and this makes its detection more comprehensive. 
In addition, \approach uses the main characteristic of spam messages, \textit{i.e.,} their need to propagate in large-scale campaigns, which can not easily be mitigated by the attackers, who, by doing so, would directly affect the efficiency of their campaigns.

We further manually examined a sample of `spam' accounts that are not detected by other approaches, to understand how \approach allows to improve the detection of spam over previous work. 

Some of the users not detected by \textsc{SpamDetector} were users aggressively advertising products, while others were bots only tweeting about a specific hashtag. 
Also these accounts have multiple bots in their friends. 
We believe that these bots were able to evade the fairly simple threat model of \textsc{SpamDetector}, but were caught by the statistical models of \approach.

Some of the missed spam accounts by \textsc{COMPA} were users who retweeted quotes to appear legitimate, but also posted spam from their blog. 
Again we found multiple bots in their friends. 
We believe that these accounts were not compromised, but rather bot accounts, and were therefore outside the threat model used by \textsc{COMPA}. 
Both \textsc{COMPA} and \textsc{SpamDetector} did not identify users posting automated content from applications. 

\textsc{BotOrNot}, on the other hand, falsely detected some benign users as bots, possibly because of their high follower-to-friend ratio and low count of tweets. It also missed some accounts that were detected by \approach, that have since been suspended by Twitter, possibly for spamming. 
These examples show that \approach is able to identify a broader category of spammers than previous systems, and is therefore more effective in fighting this problem.

\subsection{Early Spam Detection}

\approach is more effective if it can identify spam messages early on, before they are fully distributed throughout the network.
We investigated the impact of ``early detection'' on the performance of \approach. 
We implemented a simulation, where at the probabilistic table creation phase,  
for each spam observation, the propagation probabilities are only obtained for some percentage of communities, and assumes that the message has not been observed in the remaining communities.
We ran experiments for different percentages of communities, 10\%-100\%, where communities are picked randomly.  
For each percentage, we repeated the experiments three times. 

\begin{figure}[tb]
\centering
\includegraphics[width=0.40\textwidth]{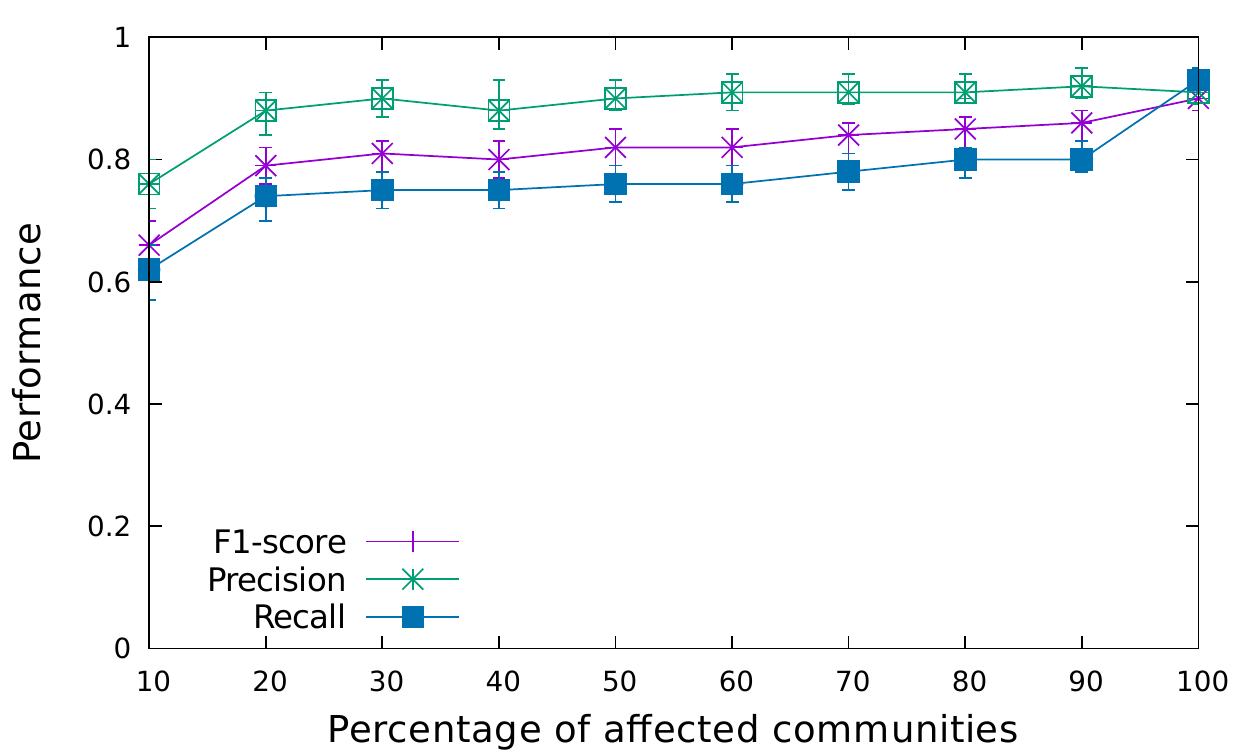}
\caption{Effective in early detection of spam messages}
\label{fig:early-detection}
\end{figure}

Figure~\ref{fig:early-detection} indicates that \approach is effective in detecting spam messages at the early stage when they have only propagated through 30\% of the communities with 90\% precision and 75\% recall.

\subsection{Adversarial Machine Learning Attacks}
\label{sec:attacks}

To evade the protection, adversaries actively manipulate data to make the classifier produce false negatives~\cite{laskov2010machine, dalvi2004adversarial, lowd2005adversarial, biggio2011bagging, Globerson:2006, Nelson:2008, Sabottke:2015}. 
These attacks will usually target two different stages of classification: 
1) at the training phase, where an adversary may attempt to mislead the classifier by ``poisoning'' its training data with carefully designed attacks~\cite{rubinstein2009antidote}, or 2) at the testing phase where an adversary may attempt to evade a deployed system by carefully manipulating attack samples~\cite{biggio2013evasion}.
These attacks are called \emph{poisoning} and \emph{evasion} attacks respectively.
We investigated the robustness of \approach against both of these adversarial settings to understand to what extent our classifier can resist the targeted attacks. 

In both attacks, an adversary attempts to make the propagation of her spam message be as similar as possible to that of a benign message in the whole network. 

For that, we assume that the adversary has the ability to create sybil accounts, establish connections with honest users and pretend to share the same interest as target communities by sending topical messages. 
As the result of these malicious activities, we assume that the adversary obtains the knowledge of: 
 1) the message counts in each of the communities of interest, and 
 2) the number of users who have posted those messages in each of those communities of interest. 
Moreover, we assume that the number of fake or compromised accounts is  equivalent to the number of users (re-)posting that specific benign message because this is one of the features used in the classifier. 

However, we assume that in non-compromised communities, the adversary is unable to control any of these variables. 
For example, let us take a network with 200 communities, where the attacker has compromised 50 communities. She observes that some similar benign messages are always posted in 30 of these communities. 
Using this information, she posts her malicious content with a similar distribution in those same 30 communities. However, since she has no control over 150 of the 200 communities, her messages most likely propagate differently than benign messages through the network.

We simulated the attacks by randomly picking some percentage of communities as \emph{compromised} by the adversary. 

\textbf{The poisoning attack} is performed during the training phase. 
At this phase, the adversary deliberately interferes with both the topic detection and community detection algorithms.   
She joins communities in order to modify the network and the community structures. 
In these now compromised communities, she posts messages resembling the propagation of benign messages in the compromised communities, therefore modifying the probability distribution of topics in these communities. 
The propagation probabilities for this spam message is actually the combination of probabilities for the chosen benign message in the \emph{compromised} communities and those of the spam message in the \emph{non-compromised} communities. 
However, since the observations in the training set are (manually) labeled, the spam messages should still be correctly labeled as spam. 
The classifier might not be able to accurately distinguish and detect either spam or benign messages because the parties of interest for benign and spam messages are partially identical.

\textbf{The evasion attack} is performed during the testing phase. The training set is not polluted while the testing set includes the instances of spam messages that attempt to resemble the probabilistic propagation of benign messages. 
In this attack, we assume that the attacker has partial knowledge about the model built on the training set, and modifies the probability distribution of her messages in compromised communities, as explained previously at the poisoning attack. 

For both attacks, in each of the neighborhoods and for each percentage of compromised communities, we ran the simulations three times and averaged all the results.  
In the case of the poisoning attack, the observations of all the spam messages in the ground-truth are modified based on the attack to represent simulated spam messages, while the observations for benign messages remain unmodified. To evaluate the attack, we performed 10-fold cross-validation on a balanced dataset generated by SMOTE algorithm~\cite{chawla2002smote}. 

In the case of the evasion attack, the testing set includes all simulated spam messages that are generated based on the attack. 
It also includes a random set of $x$ benign messages where $x$ is equal to the number of spam messages in the testing set. The training set is the entire ground-truth dataset excluding those benign messages appearing in the testing set.


\begin{figure}[tb]
\centering
\includegraphics[width=0.40\textwidth]{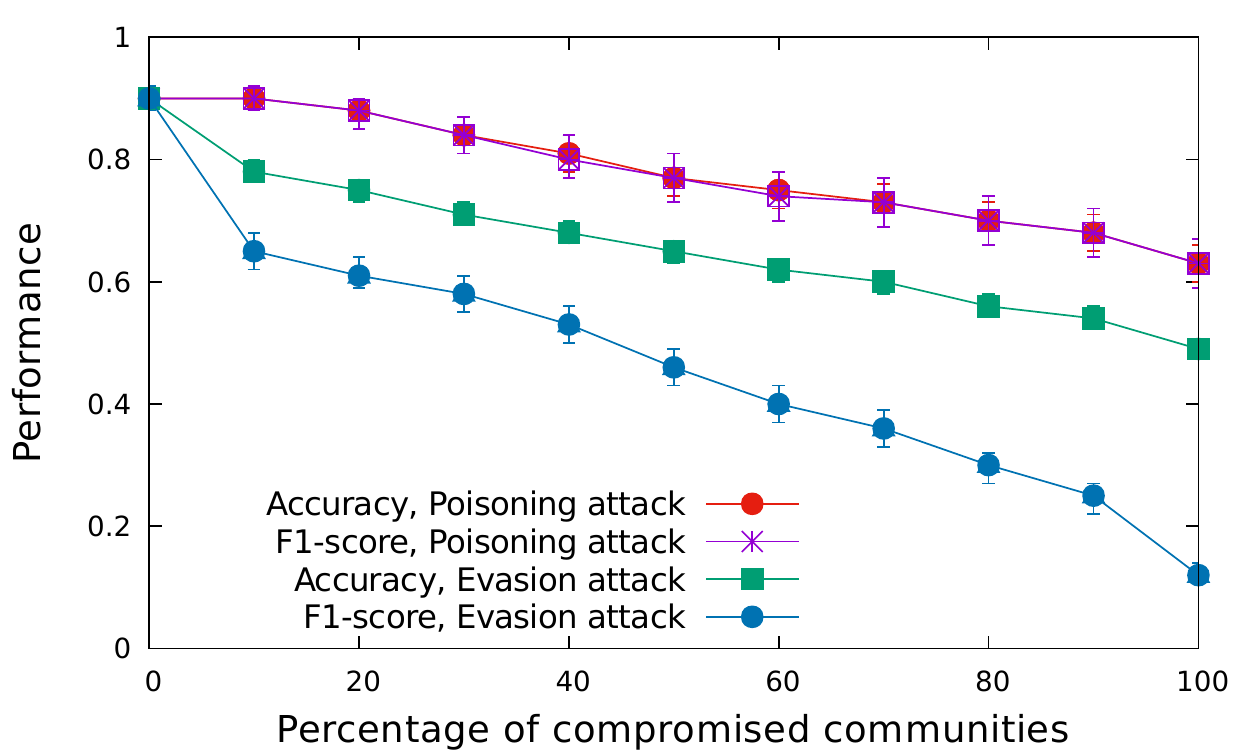}
\caption{\approach vs. adversarial machine learning attacks}
\label{fig:attacks}
\end{figure}

Figure~\ref{fig:attacks} shows the impact of these attacks on the performance of \approach when the percentage of compromised communities increases from 0 to 100. 
As it can be seen in the results, by increasing the percentage of compromised communities, the performance gradually decreases in both attacks. 
In the poisoning attack, F-1 decreases from ~90\% to ~65\%, while in the evasion attack, it decreases from ~90\% to ~15\% when the number of compromised communities increases from 0 to 100\%. 
\approach is more robust to the poisoning attack because simulated spam messages and their propagation probabilities in non-compromised communities are observed in the training set and the classifier has learned from them. 

These results suggest that the attacker needs to have a great knowledge about the network to highly impact the performance of the classifier.
For example, even if 30\% of the network is compromised, the precision and recall remain at 82\% and 87\% in the case of a poisoning attack, and at 75\% and 52\% in the case of an evasion attack.
Moreover, we argue that the poisoning attack is a more realistic attack because, in practice, the ground-truth dataset gets updated over time and the training set most probably includes spam messages generated by the adversary.


%% file: related-work.tex
\section{Related Work}
\label{related}
\approach is the first system able to detect spam on Twitter by looking at the differences in which legitimate and malicious messages propagate through the network. 
In the following, we discuss previous work in the area of spam detection on social networks. Broadly speaking we identify two types of approaches: those that look at identifying malicious messages and those that look at flagging malicious accounts. We then revise the academic literature that studied message propagation on online social networks.

\textbf{Malicious Messages Detection.}
Identifying spam messages on social networks has been extensively studied~\cite{lee2012warningbird, tan2014effect, thomas2011design, weng2013virality, xu2010toward, yardi2009detecting, ye2010measuring}. 
Yardi \textit{et al.}~\cite{yardi2009detecting} studied Twitter spammers who abuse trending topics. 
Analyzing URLs in messages is another method employed to detect malicious messages~\cite{lee2012warningbird, thomas2011design}. 
{\sc Monarch}~\cite{thomas2011design} is a system for crawling URLs spread in social network and identifying malicious messages and compromised accounts. 
Lee and Kim~\cite{lee2012warningbird} also proposed 
{\sc WarningBird}, a system that analyzes correlated redirection chains of URLs in a number of URLs posted on Twitter to identify malicious tweets. 
Some researchers have proposed offline spam analysis to identify large-scale social spam campaigns~\cite{gao2010detecting, grier2010spam}. 
They mostly apply clustering algorithms based on URL blacklisting on a complete set of messages. 
Xu \textit{et al.}~\cite{xu2010toward} presented an early warning worm detection system that monitors the behavior of users to collect suspicious worm propagation evidences.

In summary, these works concentrate on limited aspects of the spam detection problem, such as messages URL analysis, user behavior analysis, offline spam analysis, etc. 
In \approach, we provide a generic solution that detects spam messages from their propagation patterns through communities of interest, regardless of their content.

\textbf{Malicious Accounts Detection.}
Today, normal users in popular social networks are increasingly becoming the target of attackers. 
Many research works investigated this problem and proposed various solutions for this challenge~\cite{benevenuto2010detecting, boshmaf2013design, boshmaf2015integro, cai2012latent, cao2012aiding, danezis2009sybilinfer, egele2015towards, ferrara2014rise, gao2010detecting, stringhini2010detecting, wu2006topical}. 
COMPA~\cite{egele2015towards} is a system that detects compromised Twitter accounts based on their behaviors over time. The authors showed that normal users have almost stable habits over time, unlike compromised users who likely show anomalous habits. 
Liu \textit{et al.}~\cite{liu2016detecting} calculated user topics with LDA, and then employed supervised learning to identify spammers based on topics of discussion. 
Cai \textit{et al.}~\cite{cai2012latent} presented a machine learning-based platform to detect Sybil attacks in social networks. 
They split a social network into communities, and tried to identify communities that connect in an unnatural or inconsistent way with the rest of the social network. 
SybilInfer~\cite{danezis2009sybilinfer} detects compromised accounts using a Bayesian Inference approach. 
Stringhini \textit{et al.}~\cite{stringhini2010detecting} investigated spammers' behavior by creating a set of honey-profiles on popular social networks. By studying spammers' characteristics, they introduced a spam detection tool. 
Link Farming in Twitter where spammers acquire large number of follower links has been investigated by Ghosh \textit{et al.}~\cite{ghosh2012understanding}. By analyzing over 40,000 spammer accounts, they discovered that a majority of farmed links comes from a small number of legitimate and highly active users.
Wang \textit{et al.}~\cite{wang2013you} analyzed user click patterns to create user profiles and identify fake accounts using both supervised and unsupervised learning. Viswanath \textit{et al.}~\cite{viswanath2014towards} applied Principal Components Analysis (PCA) to find patterns among features extracted from spam accounts. 
Cao \textit{et al.} proposed SynchroTrap~\cite{cao2014uncovering}, a detection system that clusters malicious accounts according to their actions and the time at which they are made. 
{\sc EvilCohort}~\cite{stringhini2015evilcohort} is a system that identifies sets of social network accounts used by botnets, by looking at communities of accounts that are accessed by a common set of IP addresses.

These works aim to detect a particular type of malicious user (\textit{e.g.,} compromised accounts or fake accounts). Instead, we propose a comprehensive spam detection system independent from the type of malicious user spreading it.

\textbf{Message Propagation.}
\approach is the first system that proposes to detect spam on Twitter by looking at how legitimate and malicious messages spread on the social network.
Previous work, however, looked at message propagation on social networks for other purposes. 

Ye and Wu~\cite{ye2010measuring} studied propagation patterns of general messages and breaking news in Twitter. 
They inspected a massive number of messages collected from 700K users. 
Moreover, they evaluated different social influences by analyzing their changes over time, and how they correlate with each other. 
By analyzing Twitter hashtags, Weng \textit{et al.}~\cite{weng2013virality} showed that network communities can help predicting viral memes. In summary, the  popularity of a meme can be predicted by quantifying its early spreading pattern in terms of community concentration: the more communities a meme permeates, the more viral it is. 
Nematzadeh \textit{et al.}~\cite{nematzadeh2014optimal} demonstrated that  
strong communities with high modularity can facilitate global diffusion by enhancing local, intra-community spreading. 
Through a simulation, Mezzour \textit{et al.}~\cite{mezzour2014spam} showed how the diffusion of messages by hacked accounts differs from normal accounts. 
Similar to these works, we analyze how messages propagate in social networks. We combine this to learn parties of interest and detect spam messages.

%% file: discussion.tex
\section{Discussion}

\indent\indent \textbf{Data Collection}. 
Not having access to the whole Twitter data, including the users' data and Twitter network, imposes some constraints to our approach. For example, it may affect the quality of the communities of interest as well as the groups of similar messages. 
Nonetheless, with these limitations, \approach performed well in detecting diverse spam messages.

\textbf{Complexity and Scalability}. 
\approach is practical even at the scale of Twitter. 
Here, we discuss the complexity of the multiple phases of our approach.
The time complexity for Infomap is estimated at $O(m)$, where $m$ is the number of nodes. It can classify millions of nodes in minutes~\cite{lancichinetti2009community, aldecoa2013exploring}. 

The topic detection algorithm, LDA, has a complexity of $O(NKV)$, where $N$, $K$, and $V$ are the number of documents, topics and words in the vocabulary, respectively. 
The efficiency of LDA can be improved with the use of heuristics, \textit{e.g.,} by running it on each neighborhood independently, decreasing the number of documents by having more tweets per document, and specifying a lower number of topics.  
Recent work has also explored multiple approaches for increasing the performance of LDA~\cite{wang2009plda, nallapati2007parallelized, smyth2009asynchronous, gomes2008memory, porteous2008fast}, which can be applied to \approach.  

Identifying groups of similar messages is a string searching problem. 
\approach classifies messages based on four-gram matches. 
The length of tweets is short, and each of them only contains a few consecutive four-grams. The time complexity and memory complexity for four-gram analysis are $O(N^2M^2)$, where $N$ is the number of messages and $M$ is the maximum size of a message (\textit{i.e.,} 140 characters in the case of Twitter). However, the analysis can be optimized from $O(N^2M^2)$ to close to linear in the number of similar groups returned~\cite{kim2011text, das2007google, chum2008near}.

In our experiments, running the four-gram analysis using a commodity desktop took approximately an hour. Running LDA on 300 neighborhoods, with 15M tweets, took nine hours on a commodity desktop. Each neighborhood analysis can be run independently, and thus, can be parallelized in order to scale. 
For 500M tweets a day, which consists in the daily average on Twitter, we estimate that it would require approximately 150 machines to run our distributed analysis in two hours. 

Finally, \approach successfully detects spam messages in neighborhoods. Therefore, to increase the efficiency, \approach can be run on some partitions of networks. Furthermore, the most demanding steps are topic detection and message matching, which prepare the training data for the probabilistic model. These can be run offline and less frequently. The SVM classifier itself is highly efficient.

\textbf{Live implementation}. 
Although \approach involves many operations, testing new messages on a live system would require fewer steps. 
The actual process involved in identifying messages as spam would be to 1) establish groups of similar messages, 2) observe their propagation through parties of interest, and 3) test them on the probabilistic model that is obtained in advance. 
Twitter can collect user reports over time and use them as the ground-truth to build the probabilistic model. 
The communities of interest must be computed regularly, as the network likely evolves through time.

\textbf{Ground-truth dataset}. 
For online social networks such as Twitter and Facebook, obtaining a ground-truth dataset requires minimal effort; they already have some mechanisms in place for their users to report malicious behavior. 
Over time, following the network evolution, they can update or add to their ground-truth dataset. 
Note that our results show that \approach{} is able to detect unseen malicious content.

\textbf{Limitations}. 
Similar to other spam detection systems~\cite{egele2013compa, stringhini2012poultry}, \approach groups similar messages before applying the probabilistic model to classify spam. An attacker aware of this feature could evade the four-gram analysis used to identify messages similarity. However, in that case, the similarity analysis could be extended with more complex similarity measures~\cite{egele2013compa}. 

Our approach requires a minimum number of users and messages to form communities of interest. A malicious user with a small number of connections might be able to evade our detection. However, this goes directly against the interest of malicious users, who want to reach as many victims as possible.

%% file: conclusion.tex
\section{Conclusions and Future Work}
In this paper, we presented \approach, a novel and general approach for detecting social network spam based on its diffusion through parties of interest. 
First, we established that members of networked communities share common topics of interest distinct from other surrounding communities. 
We then built a probabilistic model that detects spam based on the dissemination of messages through communities of interest with high efficiency. 
We also showed that \approach outperforms other spam detection systems proposed in recent work, while its threat model is also more general.  
Moreover, we showed how \approach is effective in the early detection of spam messages and how it is resilient against two well-known adversarial machine learning attacks.

In this paper, we assumed that users are only member of one particular community. 
A possible future work is to explore detecting overlapping communities of interest. 
We can also explore the combination of our framework to existing systems that analyze the characteristics of user accounts or messages to detect spam.
Finally, another future work is to employ and test \approach on other online social networks.  

%% file: ms.bbl

\begin{thebibliography}{00}


\ifx \showCODEN    \undefined \def \showCODEN     #1{\unskip}     \fi
\ifx \showDOI      \undefined \def \showDOI       #1{#1}\fi
\ifx \showISBNx    \undefined \def \showISBNx     #1{\unskip}     \fi
\ifx \showISBNxiii \undefined \def \showISBNxiii  #1{\unskip}     \fi
\ifx \showISSN     \undefined \def \showISSN      #1{\unskip}     \fi
\ifx \showLCCN     \undefined \def \showLCCN      #1{\unskip}     \fi
\ifx \shownote     \undefined \def \shownote      #1{#1}          \fi
\ifx \showarticletitle \undefined \def \showarticletitle #1{#1}   \fi
\ifx \showURL      \undefined \def \showURL       {\relax}        \fi
\providecommand\bibfield[2]{#2}
\providecommand\bibinfo[2]{#2}
\providecommand\natexlab[1]{#1}
\providecommand\showeprint[2][]{arXiv:#2}

\bibitem[\protect\citeauthoryear{Ahn, Bagrow, and Lehmann}{Ahn
  et~al\mbox{.}}{2010}]%
        {ahn2010link}
\bibfield{author}{\bibinfo{person}{Yong-Yeol Ahn}, \bibinfo{person}{James~P
  Bagrow}, {and} \bibinfo{person}{Sune Lehmann}.}
  \bibinfo{year}{2010}\natexlab{}.
\newblock \showarticletitle{Link communities reveal multiscale complexity in
  networks}.
\newblock \bibinfo{journal}{{\em Nature\/}} \bibinfo{volume}{466},
  \bibinfo{number}{7307} (\bibinfo{year}{2010}), \bibinfo{pages}{761--764}.
\newblock


\bibitem[\protect\citeauthoryear{Aldecoa and Mar{\'\i}n}{Aldecoa and
  Mar{\'\i}n}{2013}]%
        {aldecoa2013exploring}
\bibfield{author}{\bibinfo{person}{Rodrigo Aldecoa} {and}
  \bibinfo{person}{Ignacio Mar{\'\i}n}.} \bibinfo{year}{2013}\natexlab{}.
\newblock \showarticletitle{Exploring the limits of community detection
  strategies in complex networks}.
\newblock \bibinfo{journal}{{\em Scientific reports\/}}  \bibinfo{volume}{3}
  (\bibinfo{year}{2013}).
\newblock


\bibitem[\protect\citeauthoryear{Benevenuto, Magno, Rodrigues, and
  Almeida}{Benevenuto et~al\mbox{.}}{2010}]%
        {benevenuto2010detecting}
\bibfield{author}{\bibinfo{person}{Fabricio Benevenuto},
  \bibinfo{person}{Gabriel Magno}, \bibinfo{person}{Tiago Rodrigues}, {and}
  \bibinfo{person}{Virgilio Almeida}.} \bibinfo{year}{2010}\natexlab{}.
\newblock \showarticletitle{Detecting spammers on twitter}. In
  \bibinfo{booktitle}{{\em Collaboration, electronic messaging, anti-abuse and
  spam conference (CEAS)}}, Vol.~\bibinfo{volume}{6}. \bibinfo{pages}{12}.
\newblock


\bibitem[\protect\citeauthoryear{Biggio, Corona, Fumera, Giacinto, and
  Roli}{Biggio et~al\mbox{.}}{2011}]%
        {biggio2011bagging}
\bibfield{author}{\bibinfo{person}{Battista Biggio}, \bibinfo{person}{Igino
  Corona}, \bibinfo{person}{Giorgio Fumera}, \bibinfo{person}{Giorgio
  Giacinto}, {and} \bibinfo{person}{Fabio Roli}.}
  \bibinfo{year}{2011}\natexlab{}.
\newblock \showarticletitle{Bagging classifiers for fighting poisoning attacks
  in adversarial classification tasks}. In \bibinfo{booktitle}{{\em
  International Workshop on Multiple Classifier Systems}}. Springer,
  \bibinfo{pages}{350--359}.
\newblock


\bibitem[\protect\citeauthoryear{Biggio, Corona, Maiorca, Nelson,
  {\v{S}}rndi{\'c}, Laskov, Giacinto, and Roli}{Biggio et~al\mbox{.}}{2013}]%
        {biggio2013evasion}
\bibfield{author}{\bibinfo{person}{Battista Biggio}, \bibinfo{person}{Igino
  Corona}, \bibinfo{person}{Davide Maiorca}, \bibinfo{person}{Blaine Nelson},
  \bibinfo{person}{Nedim {\v{S}}rndi{\'c}}, \bibinfo{person}{Pavel Laskov},
  \bibinfo{person}{Giorgio Giacinto}, {and} \bibinfo{person}{Fabio Roli}.}
  \bibinfo{year}{2013}\natexlab{}.
\newblock \showarticletitle{Evasion attacks against machine learning at test
  time}. In \bibinfo{booktitle}{{\em Joint European Conference on Machine
  Learning and Knowledge Discovery in Databases}}. Springer,
  \bibinfo{pages}{387--402}.
\newblock


\bibitem[\protect\citeauthoryear{Blei, Ng, and Jordan}{Blei
  et~al\mbox{.}}{2003}]%
        {blei2003latent}
\bibfield{author}{\bibinfo{person}{David~M Blei}, \bibinfo{person}{Andrew~Y
  Ng}, {and} \bibinfo{person}{Michael~I Jordan}.}
  \bibinfo{year}{2003}\natexlab{}.
\newblock \showarticletitle{Latent dirichlet allocation}.
\newblock \bibinfo{journal}{{\em the Journal of machine Learning research\/}}
  \bibinfo{volume}{3} (\bibinfo{year}{2003}), \bibinfo{pages}{993--1022}.
\newblock


\bibitem[\protect\citeauthoryear{Blondel, Guillaume, Lambiotte, and
  Lefebvre}{Blondel et~al\mbox{.}}{2008}]%
        {blondel2008fast}
\bibfield{author}{\bibinfo{person}{Vincent~D Blondel},
  \bibinfo{person}{Jean-Loup Guillaume}, \bibinfo{person}{Renaud Lambiotte},
  {and} \bibinfo{person}{Etienne Lefebvre}.} \bibinfo{year}{2008}\natexlab{}.
\newblock \showarticletitle{Fast unfolding of communities in large networks}.
\newblock \bibinfo{journal}{{\em Journal of Statistical Mechanics: Theory and
  Experiment\/}} \bibinfo{volume}{2008}, \bibinfo{number}{10}
  (\bibinfo{year}{2008}), \bibinfo{pages}{P10008}.
\newblock


\bibitem[\protect\citeauthoryear{Boshmaf, Logothetis, Siganos, Ler{\'\i}a,
  Lorenzo, Ripeanu, and Beznosov}{Boshmaf et~al\mbox{.}}{2015}]%
        {boshmaf2015integro}
\bibfield{author}{\bibinfo{person}{Yazan Boshmaf}, \bibinfo{person}{Dionysios
  Logothetis}, \bibinfo{person}{Georgos Siganos}, \bibinfo{person}{Jorge
  Ler{\'\i}a}, \bibinfo{person}{Jose Lorenzo}, \bibinfo{person}{Matei Ripeanu},
  {and} \bibinfo{person}{Konstantin Beznosov}.}
  \bibinfo{year}{2015}\natexlab{}.
\newblock \showarticletitle{Integro: Leveraging Victim Prediction for Robust
  Fake Account Detection in OSNs.}. In \bibinfo{booktitle}{{\em NDSS}},
  Vol.~\bibinfo{volume}{15}. \bibinfo{pages}{8--11}.
\newblock


\bibitem[\protect\citeauthoryear{Boshmaf, Muslukhov, Beznosov, and
  Ripeanu}{Boshmaf et~al\mbox{.}}{2013}]%
        {boshmaf2013design}
\bibfield{author}{\bibinfo{person}{Yazan Boshmaf}, \bibinfo{person}{Ildar
  Muslukhov}, \bibinfo{person}{Konstantin Beznosov}, {and}
  \bibinfo{person}{Matei Ripeanu}.} \bibinfo{year}{2013}\natexlab{}.
\newblock \showarticletitle{Design and analysis of a social botnet}.
\newblock \bibinfo{journal}{{\em Computer Networks\/}} \bibinfo{volume}{57},
  \bibinfo{number}{2} (\bibinfo{year}{2013}), \bibinfo{pages}{556--578}.
\newblock


\bibitem[\protect\citeauthoryear{Braga-Neto and Dougherty}{Braga-Neto and
  Dougherty}{2004}]%
        {braga2004cross}
\bibfield{author}{\bibinfo{person}{Ulisses~M Braga-Neto} {and}
  \bibinfo{person}{Edward~R Dougherty}.} \bibinfo{year}{2004}\natexlab{}.
\newblock \showarticletitle{Is cross-validation valid for small-sample
  microarray classification?}
\newblock \bibinfo{journal}{{\em Bioinformatics\/}} \bibinfo{volume}{20},
  \bibinfo{number}{3} (\bibinfo{year}{2004}), \bibinfo{pages}{374--380}.
\newblock


\bibitem[\protect\citeauthoryear{Cai and Jermaine}{Cai and Jermaine}{2012}]%
        {cai2012latent}
\bibfield{author}{\bibinfo{person}{Zhuhua Cai} {and}
  \bibinfo{person}{Christopher Jermaine}.} \bibinfo{year}{2012}\natexlab{}.
\newblock \showarticletitle{The latent community model for detecting sybil
  attacks in social networks}. In \bibinfo{booktitle}{{\em Proc. NDSS}}.
\newblock


\bibitem[\protect\citeauthoryear{Cao, Sirivianos, Yang, and Pregueiro}{Cao
  et~al\mbox{.}}{2012}]%
        {cao2012aiding}
\bibfield{author}{\bibinfo{person}{Qiang Cao}, \bibinfo{person}{Michael
  Sirivianos}, \bibinfo{person}{Xiaowei Yang}, {and} \bibinfo{person}{Tiago
  Pregueiro}.} \bibinfo{year}{2012}\natexlab{}.
\newblock \showarticletitle{Aiding the detection of fake accounts in large
  scale social online services}. In \bibinfo{booktitle}{{\em Proceedings of the
  9th USENIX conference on Networked Systems Design and Implementation}}.
  USENIX Association, \bibinfo{pages}{15--15}.
\newblock


\bibitem[\protect\citeauthoryear{Cao, Yang, Yu, and Palow}{Cao
  et~al\mbox{.}}{2014}]%
        {cao2014uncovering}
\bibfield{author}{\bibinfo{person}{Qiang Cao}, \bibinfo{person}{Xiaowei Yang},
  \bibinfo{person}{Jieqi Yu}, {and} \bibinfo{person}{Christopher Palow}.}
  \bibinfo{year}{2014}\natexlab{}.
\newblock \showarticletitle{Uncovering large groups of active malicious
  accounts in online social networks}. In \bibinfo{booktitle}{{\em Proceedings
  of the 2014 ACM SIGSAC Conference on Computer and Communications Security}}.
  ACM, \bibinfo{pages}{477--488}.
\newblock


\bibitem[\protect\citeauthoryear{Chatzakou, Kourtellis, Blackburn,
  De~Cristofaro, Stringhini, and Vakali}{Chatzakou et~al\mbox{.}}{2017}]%
        {chatzakou2017mean}
\bibfield{author}{\bibinfo{person}{Despoina Chatzakou},
  \bibinfo{person}{Nicolas Kourtellis}, \bibinfo{person}{Jeremy Blackburn},
  \bibinfo{person}{Emiliano De~Cristofaro}, \bibinfo{person}{Gianluca
  Stringhini}, {and} \bibinfo{person}{Athena Vakali}.}
  \bibinfo{year}{2017}\natexlab{}.
\newblock \showarticletitle{Mean Birds: Detecting Aggression and Bullying on
  Twitter}. In \bibinfo{booktitle}{{\em {International ACM Web Science
  Conference (WebSci)}}}.
\newblock


\bibitem[\protect\citeauthoryear{Chawla, Bowyer, Hall, and Kegelmeyer}{Chawla
  et~al\mbox{.}}{2002}]%
        {chawla2002smote}
\bibfield{author}{\bibinfo{person}{Nitesh~V. Chawla}, \bibinfo{person}{Kevin~W.
  Bowyer}, \bibinfo{person}{Lawrence~O. Hall}, {and} \bibinfo{person}{W.~Philip
  Kegelmeyer}.} \bibinfo{year}{2002}\natexlab{}.
\newblock \showarticletitle{SMOTE: synthetic minority over-sampling technique}.
\newblock \bibinfo{journal}{{\em Journal of artificial intelligence
  research\/}}  \bibinfo{volume}{16} (\bibinfo{year}{2002}),
  \bibinfo{pages}{321--357}.
\newblock


\bibitem[\protect\citeauthoryear{Cho, Myers, and Leskovec}{Cho
  et~al\mbox{.}}{2011}]%
        {Cho:2011:FMU}
\bibfield{author}{\bibinfo{person}{Eunjoon Cho}, \bibinfo{person}{Seth~A.
  Myers}, {and} \bibinfo{person}{Jure Leskovec}.}
  \bibinfo{year}{2011}\natexlab{}.
\newblock \showarticletitle{Friendship and Mobility: User Movement in
  Location-based Social Networks}. In \bibinfo{booktitle}{{\em Proceedings of
  the 17th ACM SIGKDD International Conference on Knowledge Discovery and Data
  Mining}}. \bibinfo{publisher}{ACM}, \bibinfo{address}{New York, NY, USA}.
\newblock


\bibitem[\protect\citeauthoryear{Chum, Philbin, Zisserman, et~al\mbox{.}}{Chum
  et~al\mbox{.}}{2008}]%
        {chum2008near}
\bibfield{author}{\bibinfo{person}{Ondrej Chum}, \bibinfo{person}{James
  Philbin}, \bibinfo{person}{Andrew Zisserman}, {et~al\mbox{.}}}
  \bibinfo{year}{2008}\natexlab{}.
\newblock \showarticletitle{Near Duplicate Image Detection: min-Hash and tf-idf
  Weighting.}. In \bibinfo{booktitle}{{\em BMVC}}, Vol.~\bibinfo{volume}{810}.
  \bibinfo{pages}{812--815}.
\newblock


\bibitem[\protect\citeauthoryear{Clauset, Newman, and Moore}{Clauset
  et~al\mbox{.}}{2004}]%
        {clauset2004finding}
\bibfield{author}{\bibinfo{person}{Aaron Clauset}, \bibinfo{person}{Mark~EJ
  Newman}, {and} \bibinfo{person}{Cristopher Moore}.}
  \bibinfo{year}{2004}\natexlab{}.
\newblock \showarticletitle{Finding community structure in very large
  networks}.
\newblock \bibinfo{journal}{{\em Physical review E\/}} \bibinfo{volume}{70},
  \bibinfo{number}{6} (\bibinfo{year}{2004}), \bibinfo{pages}{066111}.
\newblock


\bibitem[\protect\citeauthoryear{Cs{\'a}k{\'a}ny}{Cs{\'a}k{\'a}ny}{1981}]%
        {csakany1981homogeneity}
\bibfield{author}{\bibinfo{person}{B Cs{\'a}k{\'a}ny}.}
  \bibinfo{year}{1981}\natexlab{}.
\newblock \showarticletitle{Homogeneity and completeness}. In
  \bibinfo{booktitle}{{\em Fundamentals of Computation Theory}}. Springer,
  \bibinfo{pages}{81--89}.
\newblock


\bibitem[\protect\citeauthoryear{Culnan, McHugh, and Zubillaga}{Culnan
  et~al\mbox{.}}{2010}]%
        {culnan2010large}
\bibfield{author}{\bibinfo{person}{Mary~J Culnan}, \bibinfo{person}{Patrick~J
  McHugh}, {and} \bibinfo{person}{Jesus~I Zubillaga}.}
  \bibinfo{year}{2010}\natexlab{}.
\newblock \showarticletitle{How large US companies can use Twitter and other
  social media to gain business value}.
\newblock \bibinfo{journal}{{\em MIS Quarterly Executive\/}}
  \bibinfo{volume}{9}, \bibinfo{number}{4} (\bibinfo{year}{2010}),
  \bibinfo{pages}{243--259}.
\newblock


\bibitem[\protect\citeauthoryear{Dalvi, Domingos, Sanghai, Verma,
  et~al\mbox{.}}{Dalvi et~al\mbox{.}}{2004}]%
        {dalvi2004adversarial}
\bibfield{author}{\bibinfo{person}{Nilesh Dalvi}, \bibinfo{person}{Pedro
  Domingos}, \bibinfo{person}{Sumit Sanghai}, \bibinfo{person}{Deepak Verma},
  {et~al\mbox{.}}} \bibinfo{year}{2004}\natexlab{}.
\newblock \showarticletitle{Adversarial classification}. In
  \bibinfo{booktitle}{{\em Proceedings of the tenth ACM SIGKDD international
  conference on Knowledge discovery and data mining}}. ACM,
  \bibinfo{pages}{99--108}.
\newblock


\bibitem[\protect\citeauthoryear{Danezis and Mittal}{Danezis and
  Mittal}{2009}]%
        {danezis2009sybilinfer}
\bibfield{author}{\bibinfo{person}{George Danezis} {and}
  \bibinfo{person}{Prateek Mittal}.} \bibinfo{year}{2009}\natexlab{}.
\newblock \showarticletitle{SybilInfer: Detecting Sybil Nodes using Social
  Networks.}. In \bibinfo{booktitle}{{\em NDSS}}. San Diego, CA.
\newblock


\bibitem[\protect\citeauthoryear{Dann}{Dann}{2010}]%
        {dann2010twitter}
\bibfield{author}{\bibinfo{person}{Stephen Dann}.}
  \bibinfo{year}{2010}\natexlab{}.
\newblock \showarticletitle{Twitter content classification}.
\newblock \bibinfo{journal}{{\em First Monday\/}} \bibinfo{volume}{15},
  \bibinfo{number}{12} (\bibinfo{year}{2010}).
\newblock


\bibitem[\protect\citeauthoryear{Das, Datar, Garg, and Rajaram}{Das
  et~al\mbox{.}}{2007}]%
        {das2007google}
\bibfield{author}{\bibinfo{person}{Abhinandan~S Das}, \bibinfo{person}{Mayur
  Datar}, \bibinfo{person}{Ashutosh Garg}, {and} \bibinfo{person}{Shyam
  Rajaram}.} \bibinfo{year}{2007}\natexlab{}.
\newblock \showarticletitle{Google news personalization: scalable online
  collaborative filtering}. In \bibinfo{booktitle}{{\em Proceedings of the 16th
  international conference on World Wide Web}}. ACM, \bibinfo{pages}{271--280}.
\newblock


\bibitem[\protect\citeauthoryear{Davis, Varol, Ferrara, Flammini, and
  Menczer}{Davis et~al\mbox{.}}{2016}]%
        {davis2016botornot}
\bibfield{author}{\bibinfo{person}{Clayton~A Davis}, \bibinfo{person}{Onur
  Varol}, \bibinfo{person}{Emilio Ferrara}, \bibinfo{person}{Alessandro
  Flammini}, {and} \bibinfo{person}{Filippo Menczer}.}
  \bibinfo{year}{2016}\natexlab{}.
\newblock \showarticletitle{BotOrNot: A System to Evaluate Social Bots}.
\newblock \bibinfo{journal}{{\em arXiv preprint arXiv:1602.00975\/}}
  (\bibinfo{year}{2016}).
\newblock


\bibitem[\protect\citeauthoryear{Eaton and Mansbach}{Eaton and
  Mansbach}{2012}]%
        {eaton2012spin}
\bibfield{author}{\bibinfo{person}{Eric Eaton} {and} \bibinfo{person}{Rachael
  Mansbach}.} \bibinfo{year}{2012}\natexlab{}.
\newblock \showarticletitle{A Spin-Glass Model for Semi-Supervised Community
  Detection.}. In \bibinfo{booktitle}{{\em AAAI}}. Citeseer.
\newblock


\bibitem[\protect\citeauthoryear{Egele, Stringhini, Kruegel, and Vigna}{Egele
  et~al\mbox{.}}{2013}]%
        {egele2013compa}
\bibfield{author}{\bibinfo{person}{Manuel Egele}, \bibinfo{person}{Gianluca
  Stringhini}, \bibinfo{person}{Christopher Kruegel}, {and}
  \bibinfo{person}{Giovanni Vigna}.} \bibinfo{year}{2013}\natexlab{}.
\newblock \showarticletitle{COMPA: Detecting Compromised Accounts on Social
  Networks.}. In \bibinfo{booktitle}{{\em NDSS}}.
\newblock


\bibitem[\protect\citeauthoryear{Egele, Stringhini, Kruegel, and Vigna}{Egele
  et~al\mbox{.}}{2015}]%
        {egele2015towards}
\bibfield{author}{\bibinfo{person}{Manuel Egele}, \bibinfo{person}{Gianluca
  Stringhini}, \bibinfo{person}{Christopher Kruegel}, {and}
  \bibinfo{person}{Giovanni Vigna}.} \bibinfo{year}{2015}\natexlab{}.
\newblock \showarticletitle{Towards Detecting Compromised Accounts on Social
  Networks}.
\newblock \bibinfo{journal}{{\em {Transactions on Dependable and Secure
  Computing (TDSC)}\/}} (\bibinfo{year}{2015}).
\newblock


\bibitem[\protect\citeauthoryear{Ferrara, Varol, Davis, Menczer, and
  Flammini}{Ferrara et~al\mbox{.}}{2014}]%
        {ferrara2014rise}
\bibfield{author}{\bibinfo{person}{Emilio Ferrara}, \bibinfo{person}{Onur
  Varol}, \bibinfo{person}{Clayton Davis}, \bibinfo{person}{Filippo Menczer},
  {and} \bibinfo{person}{Alessandro Flammini}.}
  \bibinfo{year}{2014}\natexlab{}.
\newblock \showarticletitle{The rise of social bots}.
\newblock \bibinfo{journal}{{\em arXiv preprint arXiv:1407.5225\/}}
  (\bibinfo{year}{2014}).
\newblock


\bibitem[\protect\citeauthoryear{Fleizach, Voelker, and Savage}{Fleizach
  et~al\mbox{.}}{2007}]%
        {fleizach2007slicing}
\bibfield{author}{\bibinfo{person}{Chris Fleizach}, \bibinfo{person}{Geoffrey~M
  Voelker}, {and} \bibinfo{person}{Stefan Savage}.}
  \bibinfo{year}{2007}\natexlab{}.
\newblock \showarticletitle{Slicing spam with occam's razor}. In
  \bibinfo{booktitle}{{\em CEAS}}.
\newblock


\bibitem[\protect\citeauthoryear{Gao, Chen, Lee, Palsetia, and Choudhary}{Gao
  et~al\mbox{.}}{2012}]%
        {gao2012towards}
\bibfield{author}{\bibinfo{person}{Hongyu Gao}, \bibinfo{person}{Yan Chen},
  \bibinfo{person}{Kathy Lee}, \bibinfo{person}{Diana Palsetia}, {and}
  \bibinfo{person}{Alok~N Choudhary}.} \bibinfo{year}{2012}\natexlab{}.
\newblock \showarticletitle{Towards Online Spam Filtering in Social Networks.}.
  In \bibinfo{booktitle}{{\em NDSS}}.
\newblock


\bibitem[\protect\citeauthoryear{Gao, Hu, Wilson, Li, Chen, and Zhao}{Gao
  et~al\mbox{.}}{2010}]%
        {gao2010detecting}
\bibfield{author}{\bibinfo{person}{Hongyu Gao}, \bibinfo{person}{Jun Hu},
  \bibinfo{person}{Christo Wilson}, \bibinfo{person}{Zhichun Li},
  \bibinfo{person}{Yan Chen}, {and} \bibinfo{person}{Ben~Y Zhao}.}
  \bibinfo{year}{2010}\natexlab{}.
\newblock \showarticletitle{Detecting and characterizing social spam
  campaigns}. In \bibinfo{booktitle}{{\em Proceedings of the 10th ACM SIGCOMM
  conference on Internet measurement}}. ACM, \bibinfo{pages}{35--47}.
\newblock


\bibitem[\protect\citeauthoryear{Ghosh, Viswanath, Kooti, Sharma, Korlam,
  Benevenuto, Ganguly, and Gummadi}{Ghosh et~al\mbox{.}}{2012}]%
        {ghosh2012understanding}
\bibfield{author}{\bibinfo{person}{Saptarshi Ghosh}, \bibinfo{person}{Bimal
  Viswanath}, \bibinfo{person}{Farshad Kooti}, \bibinfo{person}{Naveen~Kumar
  Sharma}, \bibinfo{person}{Gautam Korlam}, \bibinfo{person}{Fabricio
  Benevenuto}, \bibinfo{person}{Niloy Ganguly}, {and}
  \bibinfo{person}{Krishna~Phani Gummadi}.} \bibinfo{year}{2012}\natexlab{}.
\newblock \showarticletitle{Understanding and combating link farming in the
  twitter social network}. In \bibinfo{booktitle}{{\em Proceedings of the 21st
  international conference on World Wide Web}}. ACM, \bibinfo{pages}{61--70}.
\newblock


\bibitem[\protect\citeauthoryear{Globerson and Roweis}{Globerson and
  Roweis}{2006}]%
        {Globerson:2006}
\bibfield{author}{\bibinfo{person}{Amir Globerson} {and} \bibinfo{person}{Sam
  Roweis}.} \bibinfo{year}{2006}\natexlab{}.
\newblock \showarticletitle{Nightmare at Test Time: Robust Learning by Feature
  Deletion}. In \bibinfo{booktitle}{{\em Proceedings of the 23rd International
  Conference on Machine Learning}} {\em (\bibinfo{series}{ICML '06})}.
  \bibinfo{publisher}{ACM}, \bibinfo{address}{New York, NY, USA}.
\newblock
\showISBNx{1-59593-383-2}


\bibitem[\protect\citeauthoryear{Gomes, Welling, and Perona}{Gomes
  et~al\mbox{.}}{2008}]%
        {gomes2008memory}
\bibfield{author}{\bibinfo{person}{Ryan Gomes}, \bibinfo{person}{Max Welling},
  {and} \bibinfo{person}{Pietro Perona}.} \bibinfo{year}{2008}\natexlab{}.
\newblock \showarticletitle{Memory bounded inference in topic models}. In
  \bibinfo{booktitle}{{\em Proceedings of the 25th international conference on
  Machine learning}}. ACM, \bibinfo{pages}{344--351}.
\newblock


\bibitem[\protect\citeauthoryear{Grier, Thomas, Paxson, and Zhang}{Grier
  et~al\mbox{.}}{2010}]%
        {grier2010spam}
\bibfield{author}{\bibinfo{person}{Chris Grier}, \bibinfo{person}{Kurt Thomas},
  \bibinfo{person}{Vern Paxson}, {and} \bibinfo{person}{Michael Zhang}.}
  \bibinfo{year}{2010}\natexlab{}.
\newblock \showarticletitle{@ spam: the underground on 140 characters or less}.
  In \bibinfo{booktitle}{{\em Proceedings of the 17th ACM conference on
  Computer and communications security}}. ACM, \bibinfo{pages}{27--37}.
\newblock


\bibitem[\protect\citeauthoryear{Hong and Davison}{Hong and Davison}{2010}]%
        {hong2010empirical}
\bibfield{author}{\bibinfo{person}{Liangjie Hong} {and}
  \bibinfo{person}{Brian~D Davison}.} \bibinfo{year}{2010}\natexlab{}.
\newblock \showarticletitle{Empirical study of topic modeling in twitter}. In
  \bibinfo{booktitle}{{\em Proceedings of the first workshop on social media
  analytics}}. ACM, \bibinfo{pages}{80--88}.
\newblock


\bibitem[\protect\citeauthoryear{Jagatic, Johnson, Jakobsson, and
  Menczer}{Jagatic et~al\mbox{.}}{2007}]%
        {jagatic2007social}
\bibfield{author}{\bibinfo{person}{Tom~N Jagatic}, \bibinfo{person}{Nathaniel~A
  Johnson}, \bibinfo{person}{Markus Jakobsson}, {and} \bibinfo{person}{Filippo
  Menczer}.} \bibinfo{year}{2007}\natexlab{}.
\newblock \showarticletitle{Social phishing}.
\newblock \bibinfo{journal}{{\it Commun. ACM}} \bibinfo{volume}{50},
  \bibinfo{number}{10} (\bibinfo{year}{2007}), \bibinfo{pages}{94--100}.
\newblock


\bibitem[\protect\citeauthoryear{Java, Song, Finin, and Tseng}{Java
  et~al\mbox{.}}{2007}]%
        {java2007we}
\bibfield{author}{\bibinfo{person}{Akshay Java}, \bibinfo{person}{Xiaodan
  Song}, \bibinfo{person}{Tim Finin}, {and} \bibinfo{person}{Belle Tseng}.}
  \bibinfo{year}{2007}\natexlab{}.
\newblock \showarticletitle{Why we twitter: understanding microblogging usage
  and communities}. In \bibinfo{booktitle}{{\em Proceedings of the 9th WebKDD
  and 1st SNA-KDD 2007 workshop on Web mining and social network analysis}}.
  ACM, \bibinfo{pages}{56--65}.
\newblock


\bibitem[\protect\citeauthoryear{Kim and Shim}{Kim and Shim}{2011}]%
        {kim2011text}
\bibfield{author}{\bibinfo{person}{Chulyun Kim} {and} \bibinfo{person}{Kyuseok
  Shim}.} \bibinfo{year}{2011}\natexlab{}.
\newblock \showarticletitle{Text: Automatic template extraction from
  heterogeneous web pages}.
\newblock \bibinfo{journal}{{\em IEEE Transactions on knowledge and data
  Engineering\/}} \bibinfo{volume}{23}, \bibinfo{number}{4}
  (\bibinfo{year}{2011}), \bibinfo{pages}{612--626}.
\newblock


\bibitem[\protect\citeauthoryear{Kohavi}{Kohavi}{1995}]%
        {kohavi1995study}
\bibfield{author}{\bibinfo{person}{Ron Kohavi}.}
  \bibinfo{year}{1995}\natexlab{}.
\newblock \showarticletitle{A Study of Cross-validation and Bootstrap for
  Accuracy Estimation and Model Selection}. In \bibinfo{booktitle}{{\em
  Proceedings of the 14th International Joint Conference on Artificial
  Intelligence - Volume 2}} {\em (\bibinfo{series}{IJCAI'95})}.
  \bibinfo{publisher}{Morgan Kaufmann Publishers Inc.}
\newblock


\bibitem[\protect\citeauthoryear{Kwak, Lee, Park, and Moon}{Kwak
  et~al\mbox{.}}{2010}]%
        {Kwak:2010}
\bibfield{author}{\bibinfo{person}{Haewoon Kwak}, \bibinfo{person}{Changhyun
  Lee}, \bibinfo{person}{Hosung Park}, {and} \bibinfo{person}{Sue Moon}.}
  \bibinfo{year}{2010}\natexlab{}.
\newblock \showarticletitle{What is Twitter, a Social Network or a News
  Media?}. In \bibinfo{booktitle}{{\em Proceedings of the 19th International
  Conference on World Wide Web}}. \bibinfo{publisher}{ACM}.
\newblock


\bibitem[\protect\citeauthoryear{Lancichinetti and Fortunato}{Lancichinetti and
  Fortunato}{2009a}]%
        {lancichinetti2009benchmarks}
\bibfield{author}{\bibinfo{person}{Andrea Lancichinetti} {and}
  \bibinfo{person}{Santo Fortunato}.} \bibinfo{year}{2009}\natexlab{a}.
\newblock \showarticletitle{Benchmarks for testing community detection
  algorithms on directed and weighted graphs with overlapping communities}.
\newblock \bibinfo{journal}{{\em Physical Review E\/}} \bibinfo{volume}{80},
  \bibinfo{number}{1} (\bibinfo{year}{2009}), \bibinfo{pages}{016118}.
\newblock


\bibitem[\protect\citeauthoryear{Lancichinetti and Fortunato}{Lancichinetti and
  Fortunato}{2009b}]%
        {lancichinetti2009community}
\bibfield{author}{\bibinfo{person}{Andrea Lancichinetti} {and}
  \bibinfo{person}{Santo Fortunato}.} \bibinfo{year}{2009}\natexlab{b}.
\newblock \showarticletitle{Community detection algorithms: a comparative
  analysis}.
\newblock \bibinfo{journal}{{\em Physical review E\/}} \bibinfo{volume}{80},
  \bibinfo{number}{5} (\bibinfo{year}{2009}), \bibinfo{pages}{056117}.
\newblock


\bibitem[\protect\citeauthoryear{Lancichinetti, Sirer, Wang, Acuna, Körding,
  and Amaral}{Lancichinetti et~al\mbox{.}}{2015}]%
        {Lancichinetti2015}
\bibfield{author}{\bibinfo{person}{A Lancichinetti}, \bibinfo{person}{MI
  Sirer}, \bibinfo{person}{JX Wang}, \bibinfo{person}{D Acuna},
  \bibinfo{person}{K Körding}, {and} \bibinfo{person}{LAN Amaral}.}
  \bibinfo{year}{2015}\natexlab{}.
\newblock \showarticletitle{High-reproducibility and high-accuracy method for
  automated topic classification}.
\newblock \bibinfo{journal}{{\em Physical Review X\/}}  \bibinfo{volume}{5}
  (\bibinfo{date}{29 JAN} \bibinfo{year}{2015}), \bibinfo{pages}{011007}.
\newblock
\showDOI{%
\url{https://doi.org/10.1103/PhysRevX.5.011007}}


\bibitem[\protect\citeauthoryear{Laskov and Lippmann}{Laskov and
  Lippmann}{2010}]%
        {laskov2010machine}
\bibfield{author}{\bibinfo{person}{Pavel Laskov} {and} \bibinfo{person}{Richard
  Lippmann}.} \bibinfo{year}{2010}\natexlab{}.
\newblock \showarticletitle{Machine learning in adversarial environments}.
\newblock \bibinfo{journal}{{\em Machine learning\/}} \bibinfo{volume}{81},
  \bibinfo{number}{2} (\bibinfo{year}{2010}), \bibinfo{pages}{115--119}.
\newblock


\bibitem[\protect\citeauthoryear{Lee and Kim}{Lee and Kim}{2012}]%
        {lee2012warningbird}
\bibfield{author}{\bibinfo{person}{Sangho Lee} {and} \bibinfo{person}{Jong
  Kim}.} \bibinfo{year}{2012}\natexlab{}.
\newblock \showarticletitle{WarningBird: Detecting Suspicious URLs in Twitter
  Stream.}. In \bibinfo{booktitle}{{\em NDSS}}.
\newblock


\bibitem[\protect\citeauthoryear{Lerman and Ghosh}{Lerman and Ghosh}{2010}]%
        {lerman2010information}
\bibfield{author}{\bibinfo{person}{Kristina Lerman} {and} \bibinfo{person}{Rumi
  Ghosh}.} \bibinfo{year}{2010}\natexlab{}.
\newblock \showarticletitle{Information contagion: An empirical study of the
  spread of news on Digg and Twitter social networks.}
\newblock \bibinfo{journal}{{\em ICWSM\/}}  \bibinfo{volume}{10}
  (\bibinfo{year}{2010}), \bibinfo{pages}{90--97}.
\newblock


\bibitem[\protect\citeauthoryear{Lim and Datta}{Lim and Datta}{2012}]%
        {lim2012finding}
\bibfield{author}{\bibinfo{person}{Kwan~Hui Lim} {and} \bibinfo{person}{Amitava
  Datta}.} \bibinfo{year}{2012}\natexlab{}.
\newblock \showarticletitle{Finding twitter communities with common interests
  using following links of celebrities}. In \bibinfo{booktitle}{{\em
  Proceedings of the 3rd international workshop on Modeling social media}}.
  ACM, \bibinfo{pages}{25--32}.
\newblock


\bibitem[\protect\citeauthoryear{Liu, Gao, Wright, and Mittal}{Liu
  et~al\mbox{.}}{2015}]%
        {liu2015exploiting}
\bibfield{author}{\bibinfo{person}{Changchang Liu}, \bibinfo{person}{Peng Gao},
  \bibinfo{person}{Matthew Wright}, {and} \bibinfo{person}{Prateek Mittal}.}
  \bibinfo{year}{2015}\natexlab{}.
\newblock \showarticletitle{Exploiting Temporal Dynamics in Sybil Defenses}. In
  \bibinfo{booktitle}{{\em ACM SIGSAC Conference on Computer and Communications
  Security (CCS)}}.
\newblock


\bibitem[\protect\citeauthoryear{Liu, Lu, Luo, Zhang, Itti, and Lu}{Liu
  et~al\mbox{.}}{2016}]%
        {liu2016detecting}
\bibfield{author}{\bibinfo{person}{Linqing Liu}, \bibinfo{person}{Yao Lu},
  \bibinfo{person}{Ye Luo}, \bibinfo{person}{Renxian Zhang},
  \bibinfo{person}{Laurent Itti}, {and} \bibinfo{person}{Jianwei Lu}.}
  \bibinfo{year}{2016}\natexlab{}.
\newblock \showarticletitle{Detecting" Smart" Spammers On Social Network: A
  Topic Model Approach}.
\newblock \bibinfo{journal}{{\em arXiv preprint arXiv:1604.08504\/}}
  (\bibinfo{year}{2016}).
\newblock


\bibitem[\protect\citeauthoryear{Lowd and Meek}{Lowd and Meek}{2005}]%
        {lowd2005adversarial}
\bibfield{author}{\bibinfo{person}{Daniel Lowd} {and}
  \bibinfo{person}{Christopher Meek}.} \bibinfo{year}{2005}\natexlab{}.
\newblock \showarticletitle{Adversarial learning}. In \bibinfo{booktitle}{{\em
  Proceedings of the eleventh ACM SIGKDD international conference on Knowledge
  discovery in data mining}}. ACM, \bibinfo{pages}{641--647}.
\newblock


\bibitem[\protect\citeauthoryear{McCallum}{McCallum}{2002}]%
        {mccallum2002mallet}
\bibfield{author}{\bibinfo{person}{Andrew~K McCallum}.}
  \bibinfo{year}{2002}\natexlab{}.
\newblock \bibinfo{title}{$\{$MALLET: A Machine Learning for Language
  Toolkit$\}$}.
\newblock \bibinfo{howpublished}{\url{http://mallet.cs.umass.edu/}}.
  (\bibinfo{year}{2002}).
\newblock


\bibitem[\protect\citeauthoryear{McPherson, Smith-Lovin, and Cook}{McPherson
  et~al\mbox{.}}{2001}]%
        {mcpherson2001birds}
\bibfield{author}{\bibinfo{person}{Miller McPherson}, \bibinfo{person}{Lynn
  Smith-Lovin}, {and} \bibinfo{person}{James~M Cook}.}
  \bibinfo{year}{2001}\natexlab{}.
\newblock \showarticletitle{Birds of a feather: Homophily in social networks}.
\newblock \bibinfo{journal}{{\em Annual review of sociology\/}}
  (\bibinfo{year}{2001}), \bibinfo{pages}{415--444}.
\newblock


\bibitem[\protect\citeauthoryear{Mezzour and Carley}{Mezzour and
  Carley}{2014}]%
        {mezzour2014spam}
\bibfield{author}{\bibinfo{person}{Ghita Mezzour} {and}
  \bibinfo{person}{Kathleen~M Carley}.} \bibinfo{year}{2014}\natexlab{}.
\newblock \showarticletitle{Spam diffusion in a social network initiated by
  hacked e--mail accounts}.
\newblock \bibinfo{journal}{{\em International Journal of Security and
  Networks\/}} \bibinfo{volume}{9}, \bibinfo{number}{3} (\bibinfo{year}{2014}),
  \bibinfo{pages}{144--153}.
\newblock


\bibitem[\protect\citeauthoryear{Mucha, Richardson, Macon, Porter, and
  Onnela}{Mucha et~al\mbox{.}}{2010}]%
        {mucha2010community}
\bibfield{author}{\bibinfo{person}{Peter~J Mucha}, \bibinfo{person}{Thomas
  Richardson}, \bibinfo{person}{Kevin Macon}, \bibinfo{person}{Mason~A Porter},
  {and} \bibinfo{person}{Jukka-Pekka Onnela}.} \bibinfo{year}{2010}\natexlab{}.
\newblock \showarticletitle{Community structure in time-dependent, multiscale,
  and multiplex networks}.
\newblock \bibinfo{journal}{{\em science\/}} \bibinfo{volume}{328},
  \bibinfo{number}{5980} (\bibinfo{year}{2010}), \bibinfo{pages}{876--878}.
\newblock


\bibitem[\protect\citeauthoryear{Nallapati, Cohen, and Lafferty}{Nallapati
  et~al\mbox{.}}{2007}]%
        {nallapati2007parallelized}
\bibfield{author}{\bibinfo{person}{Ramesh Nallapati}, \bibinfo{person}{William
  Cohen}, {and} \bibinfo{person}{John Lafferty}.}
  \bibinfo{year}{2007}\natexlab{}.
\newblock \showarticletitle{Parallelized variational EM for latent Dirichlet
  allocation: An experimental evaluation of speed and scalability}. In
  \bibinfo{booktitle}{{\em Seventh IEEE International Conference on Data Mining
  Workshops (ICDMW 2007)}}. IEEE, \bibinfo{pages}{349--354}.
\newblock


\bibitem[\protect\citeauthoryear{Nelson, Barreno, Chi, Joseph, Rubinstein,
  Saini, Sutton, Tygar, and Xia}{Nelson et~al\mbox{.}}{2008}]%
        {Nelson:2008}
\bibfield{author}{\bibinfo{person}{Blaine Nelson}, \bibinfo{person}{Marco
  Barreno}, \bibinfo{person}{Fuching~Jack Chi}, \bibinfo{person}{Anthony~D.
  Joseph}, \bibinfo{person}{Benjamin I.~P. Rubinstein}, \bibinfo{person}{Udam
  Saini}, \bibinfo{person}{Charles Sutton}, \bibinfo{person}{J.~D. Tygar},
  {and} \bibinfo{person}{Kai Xia}.} \bibinfo{year}{2008}\natexlab{}.
\newblock \showarticletitle{Exploiting Machine Learning to Subvert Your Spam
  Filter}. In \bibinfo{booktitle}{{\em Proceedings of the 1st Usenix Workshop
  on Large-Scale Exploits and Emergent Threats}} {\em
  (\bibinfo{series}{LEET'08})}. \bibinfo{publisher}{USENIX Association},
  \bibinfo{address}{Berkeley, CA, USA}.
\newblock


\bibitem[\protect\citeauthoryear{Nematzadeh, Ferrara, Flammini, and
  Ahn}{Nematzadeh et~al\mbox{.}}{2014}]%
        {nematzadeh2014optimal}
\bibfield{author}{\bibinfo{person}{Azadeh Nematzadeh}, \bibinfo{person}{Emilio
  Ferrara}, \bibinfo{person}{Alessandro Flammini}, {and}
  \bibinfo{person}{Yong-Yeol Ahn}.} \bibinfo{year}{2014}\natexlab{}.
\newblock \showarticletitle{Optimal network modularity for information
  diffusion}.
\newblock \bibinfo{journal}{{\em Physical review letters\/}}
  \bibinfo{volume}{113}, \bibinfo{number}{8} (\bibinfo{year}{2014}),
  \bibinfo{pages}{088701}.
\newblock


\bibitem[\protect\citeauthoryear{Newman}{Newman}{2006}]%
        {newman2006modularity}
\bibfield{author}{\bibinfo{person}{Mark~EJ Newman}.}
  \bibinfo{year}{2006}\natexlab{}.
\newblock \showarticletitle{Modularity and community structure in networks}.
\newblock \bibinfo{journal}{{\em Proceedings of the National Academy of
  Sciences\/}} \bibinfo{volume}{103}, \bibinfo{number}{23}
  (\bibinfo{year}{2006}), \bibinfo{pages}{8577--8582}.
\newblock


\bibitem[\protect\citeauthoryear{Newman and Park}{Newman and Park}{2003}]%
        {newman2003social}
\bibfield{author}{\bibinfo{person}{Mark~EJ Newman} {and}
  \bibinfo{person}{Juyong Park}.} \bibinfo{year}{2003}\natexlab{}.
\newblock \showarticletitle{Why social networks are different from other types
  of networks}.
\newblock \bibinfo{journal}{{\em Physical Review E\/}} \bibinfo{volume}{68},
  \bibinfo{number}{3} (\bibinfo{year}{2003}), \bibinfo{pages}{036122}.
\newblock


\bibitem[\protect\citeauthoryear{Palla, Der\'{e}nyi, Farkas, and Vicsek}{Palla
  et~al\mbox{.}}{2005}]%
        {Palla05}
\bibfield{author}{\bibinfo{person}{Gergely Palla}, \bibinfo{person}{Imre
  Der\'{e}nyi}, \bibinfo{person}{Ill\'{e}s Farkas}, {and}
  \bibinfo{person}{Tam\'{a}s Vicsek}.} \bibinfo{year}{2005}\natexlab{}.
\newblock \showarticletitle{Uncovering the overlapping community structure of
  complex networks in nature and society}.
\newblock \bibinfo{journal}{{\em Nature\/}} \bibinfo{volume}{435},
  \bibinfo{number}{7043} (\bibinfo{date}{June} \bibinfo{year}{2005}),
  \bibinfo{pages}{814--818}.
\newblock
\showISSN{0028-0836}


\bibitem[\protect\citeauthoryear{Perry and Wolfe}{Perry and Wolfe}{2012}]%
        {perry2012null}
\bibfield{author}{\bibinfo{person}{Patrick~O Perry} {and}
  \bibinfo{person}{Patrick~J Wolfe}.} \bibinfo{year}{2012}\natexlab{}.
\newblock \showarticletitle{Null models for network data}.
\newblock \bibinfo{journal}{{\em arXiv preprint arXiv:1201.5871\/}}
  (\bibinfo{year}{2012}).
\newblock


\bibitem[\protect\citeauthoryear{Pons and Latapy}{Pons and Latapy}{2005}]%
        {pons2005computing}
\bibfield{author}{\bibinfo{person}{Pascal Pons} {and} \bibinfo{person}{Matthieu
  Latapy}.} \bibinfo{year}{2005}\natexlab{}.
\newblock \showarticletitle{Computing communities in large networks using
  random walks}.
\newblock In \bibinfo{booktitle}{{\em Computer and Information Sciences-ISCIS
  2005}}. \bibinfo{publisher}{Springer}, \bibinfo{pages}{284--293}.
\newblock


\bibitem[\protect\citeauthoryear{Porteous, Newman, Ihler, Asuncion, Smyth, and
  Welling}{Porteous et~al\mbox{.}}{2008}]%
        {porteous2008fast}
\bibfield{author}{\bibinfo{person}{Ian Porteous}, \bibinfo{person}{David
  Newman}, \bibinfo{person}{Alexander Ihler}, \bibinfo{person}{Arthur
  Asuncion}, \bibinfo{person}{Padhraic Smyth}, {and} \bibinfo{person}{Max
  Welling}.} \bibinfo{year}{2008}\natexlab{}.
\newblock \showarticletitle{Fast collapsed gibbs sampling for latent dirichlet
  allocation}. In \bibinfo{booktitle}{{\em Proceedings of the 14th ACM SIGKDD
  international conference on Knowledge discovery and data mining}}. ACM,
  \bibinfo{pages}{569--577}.
\newblock


\bibitem[\protect\citeauthoryear{Refaeilzadeh, Tang, and Liu}{Refaeilzadeh
  et~al\mbox{.}}{2009}]%
        {refaeilzadeh2009cross}
\bibfield{author}{\bibinfo{person}{Payam Refaeilzadeh}, \bibinfo{person}{Lei
  Tang}, {and} \bibinfo{person}{Huan Liu}.} \bibinfo{year}{2009}\natexlab{}.
\newblock \showarticletitle{Cross-validation}.
\newblock In \bibinfo{booktitle}{{\em Encyclopedia of database systems}}.
  \bibinfo{publisher}{Springer}, \bibinfo{pages}{532--538}.
\newblock


\bibitem[\protect\citeauthoryear{Rosa, Shah, Lin, Gershman, and
  Frederking}{Rosa et~al\mbox{.}}{2011}]%
        {rosa2011topical}
\bibfield{author}{\bibinfo{person}{Kevin~Dela Rosa}, \bibinfo{person}{Rushin
  Shah}, \bibinfo{person}{Bo Lin}, \bibinfo{person}{Anatole Gershman}, {and}
  \bibinfo{person}{Robert Frederking}.} \bibinfo{year}{2011}\natexlab{}.
\newblock \showarticletitle{Topical clustering of tweets}.
\newblock \bibinfo{journal}{{\em Proceedings of the ACM SIGIR: SWSM\/}}
  (\bibinfo{year}{2011}).
\newblock


\bibitem[\protect\citeauthoryear{Rosenberg and Hirschberg}{Rosenberg and
  Hirschberg}{2007}]%
        {rosenberg2007v}
\bibfield{author}{\bibinfo{person}{Andrew Rosenberg} {and}
  \bibinfo{person}{Julia Hirschberg}.} \bibinfo{year}{2007}\natexlab{}.
\newblock \showarticletitle{V-Measure: A Conditional Entropy-Based External
  Cluster Evaluation Measure.}. In \bibinfo{booktitle}{{\em EMNLP-CoNLL}},
  Vol.~\bibinfo{volume}{7}. \bibinfo{pages}{410--420}.
\newblock


\bibitem[\protect\citeauthoryear{Rosvall and Bergstrom}{Rosvall and
  Bergstrom}{2008}]%
        {rosvall2008maps}
\bibfield{author}{\bibinfo{person}{Martin Rosvall} {and}
  \bibinfo{person}{Carl~T Bergstrom}.} \bibinfo{year}{2008}\natexlab{}.
\newblock \showarticletitle{Maps of random walks on complex networks reveal
  community structure}.
\newblock \bibinfo{journal}{{\em Proceedings of the National Academy of
  Sciences\/}} \bibinfo{volume}{105}, \bibinfo{number}{4}
  (\bibinfo{year}{2008}), \bibinfo{pages}{1118--1123}.
\newblock


\bibitem[\protect\citeauthoryear{Rubinstein, Nelson, Huang, Joseph, Lau, Rao,
  Taft, and Tygar}{Rubinstein et~al\mbox{.}}{2009}]%
        {rubinstein2009antidote}
\bibfield{author}{\bibinfo{person}{Benjamin~IP Rubinstein},
  \bibinfo{person}{Blaine Nelson}, \bibinfo{person}{Ling Huang},
  \bibinfo{person}{Anthony~D Joseph}, \bibinfo{person}{Shing-hon Lau},
  \bibinfo{person}{Satish Rao}, \bibinfo{person}{Nina Taft}, {and}
  \bibinfo{person}{JD Tygar}.} \bibinfo{year}{2009}\natexlab{}.
\newblock \showarticletitle{Antidote: understanding and defending against
  poisoning of anomaly detectors}. In \bibinfo{booktitle}{{\em Proceedings of
  the 9th ACM SIGCOMM conference on Internet measurement conference}}. ACM,
  \bibinfo{pages}{1--14}.
\newblock


\bibitem[\protect\citeauthoryear{Sabottke, Suciu, and Dumitras}{Sabottke
  et~al\mbox{.}}{2015}]%
        {Sabottke:2015}
\bibfield{author}{\bibinfo{person}{Carl Sabottke}, \bibinfo{person}{Octavian
  Suciu}, {and} \bibinfo{person}{Tudor Dumitras}.}
  \bibinfo{year}{2015}\natexlab{}.
\newblock \showarticletitle{Vulnerability Disclosure in the Age of Social
  Media: Exploiting Twitter for Predicting Real-world Exploits}. In
  \bibinfo{booktitle}{{\em Proceedings of the 24th USENIX Conference on
  Security Symposium}} {\em (\bibinfo{series}{SEC'15})}.
  \bibinfo{publisher}{USENIX Association}, \bibinfo{address}{Berkeley, CA,
  USA}, \bibinfo{pages}{1041--1056}.
\newblock
\showISBNx{978-1-931971-232}
\showURL{%
\url{http://dl.acm.org/citation.cfm?id=2831143.2831209}}


\bibitem[\protect\citeauthoryear{Seidman}{Seidman}{1983}]%
        {seidman1983network}
\bibfield{author}{\bibinfo{person}{Stephen~B Seidman}.}
  \bibinfo{year}{1983}\natexlab{}.
\newblock \showarticletitle{Network structure and minimum degree}.
\newblock \bibinfo{journal}{{\em Social networks\/}} \bibinfo{volume}{5},
  \bibinfo{number}{3} (\bibinfo{year}{1983}), \bibinfo{pages}{269--287}.
\newblock


\bibitem[\protect\citeauthoryear{Smyth, Welling, and Asuncion}{Smyth
  et~al\mbox{.}}{2009}]%
        {smyth2009asynchronous}
\bibfield{author}{\bibinfo{person}{Padhraic Smyth}, \bibinfo{person}{Max
  Welling}, {and} \bibinfo{person}{Arthur~U Asuncion}.}
  \bibinfo{year}{2009}\natexlab{}.
\newblock \showarticletitle{Asynchronous distributed learning of topic models}.
  In \bibinfo{booktitle}{{\em Advances in Neural Information Processing
  Systems}}. \bibinfo{pages}{81--88}.
\newblock


\bibitem[\protect\citeauthoryear{Stringhini, Egele, Kruegel, and
  Vigna}{Stringhini et~al\mbox{.}}{2012}]%
        {stringhini2012poultry}
\bibfield{author}{\bibinfo{person}{G. Stringhini}, \bibinfo{person}{M. Egele},
  \bibinfo{person}{C. Kruegel}, {and} \bibinfo{person}{G. Vigna}.}
  \bibinfo{year}{2012}\natexlab{}.
\newblock \showarticletitle{{Poultry Markets: On the Underground Economy of
  Twitter Followers}}. In \bibinfo{booktitle}{{\em Proceedings of the Workshop
  on Online Social Network (WOSN)}}. ACM, \bibinfo{address}{Helsinki, Finland}.
\newblock


\bibitem[\protect\citeauthoryear{Stringhini, Kruegel, and Vigna}{Stringhini
  et~al\mbox{.}}{2010}]%
        {stringhini2010detecting}
\bibfield{author}{\bibinfo{person}{Gianluca Stringhini},
  \bibinfo{person}{Christopher Kruegel}, {and} \bibinfo{person}{Giovanni
  Vigna}.} \bibinfo{year}{2010}\natexlab{}.
\newblock \showarticletitle{Detecting spammers on social networks}. In
  \bibinfo{booktitle}{{\em Proceedings of the 26th Annual Computer Security
  Applications Conference}}. ACM, \bibinfo{pages}{1--9}.
\newblock


\bibitem[\protect\citeauthoryear{Stringhini, Mourlanne, Jacob, Egele, Kruegel,
  and Vigna}{Stringhini et~al\mbox{.}}{2015}]%
        {stringhini2015evilcohort}
\bibfield{author}{\bibinfo{person}{Gianluca Stringhini},
  \bibinfo{person}{Pierre Mourlanne}, \bibinfo{person}{Gregoire Jacob},
  \bibinfo{person}{Manuel Egele}, \bibinfo{person}{Christopher Kruegel}, {and}
  \bibinfo{person}{Giovanni Vigna}.} \bibinfo{year}{2015}\natexlab{}.
\newblock \showarticletitle{Evilcohort: detecting communities of malicious
  accounts on online services}. In \bibinfo{booktitle}{{\em USENIX Security
  Symposium}}.
\newblock


\bibitem[\protect\citeauthoryear{Su and Wu}{Su and Wu}{2013}]%
        {su2013null}
\bibfield{author}{\bibinfo{person}{Jessica Su} {and} \bibinfo{person}{Sen Wu}.}
  \bibinfo{year}{2013}\natexlab{}.
\newblock \bibinfo{title}{Null Models For Social Networks}.
\newblock
  \bibinfo{howpublished}{\url{http://snap.stanford.edu/class/cs224w-2013/projects2013/cs224w-003-final.pdf}}.
    (\bibinfo{year}{2013}).
\newblock


\bibitem[\protect\citeauthoryear{Tan, Lee, and Pang}{Tan et~al\mbox{.}}{2014}]%
        {tan2014effect}
\bibfield{author}{\bibinfo{person}{Chenhao Tan}, \bibinfo{person}{Lillian Lee},
  {and} \bibinfo{person}{Bo Pang}.} \bibinfo{year}{2014}\natexlab{}.
\newblock \showarticletitle{The effect of wording on message propagation:
  Topic-and author-controlled natural experiments on Twitter}.
\newblock \bibinfo{journal}{{\em arXiv preprint arXiv:1405.1438\/}}
  (\bibinfo{year}{2014}).
\newblock


\bibitem[\protect\citeauthoryear{Tang and Liu}{Tang and Liu}{2010}]%
        {tang2010community}
\bibfield{author}{\bibinfo{person}{Lei Tang} {and} \bibinfo{person}{Huan Liu}.}
  \bibinfo{year}{2010}\natexlab{}.
\newblock \showarticletitle{Community detection and mining in social media}.
\newblock \bibinfo{journal}{{\em Synthesis Lectures on Data Mining and
  Knowledge Discovery\/}} \bibinfo{volume}{2}, \bibinfo{number}{1}
  (\bibinfo{year}{2010}).
\newblock


\bibitem[\protect\citeauthoryear{Thomas, Grier, Ma, Paxson, and Song}{Thomas
  et~al\mbox{.}}{2011a}]%
        {thomas2011design}
\bibfield{author}{\bibinfo{person}{Kurt Thomas}, \bibinfo{person}{Chris Grier},
  \bibinfo{person}{Justin Ma}, \bibinfo{person}{Vern Paxson}, {and}
  \bibinfo{person}{Dawn Song}.} \bibinfo{year}{2011}\natexlab{a}.
\newblock \showarticletitle{Design and evaluation of a real-time url spam
  filtering service}. In \bibinfo{booktitle}{{\em Security and Privacy (SP),
  2011 IEEE Symposium on}}. IEEE, \bibinfo{pages}{447--462}.
\newblock


\bibitem[\protect\citeauthoryear{Thomas, Grier, Song, and Paxson}{Thomas
  et~al\mbox{.}}{2011b}]%
        {thomas2011suspended}
\bibfield{author}{\bibinfo{person}{Kurt Thomas}, \bibinfo{person}{Chris Grier},
  \bibinfo{person}{Dawn Song}, {and} \bibinfo{person}{Vern Paxson}.}
  \bibinfo{year}{2011}\natexlab{b}.
\newblock \showarticletitle{Suspended accounts in retrospect: an analysis of
  twitter spam}. In \bibinfo{booktitle}{{\em Proceedings of the 2011 ACM
  SIGCOMM conference on Internet measurement conference}}. ACM,
  \bibinfo{pages}{243--258}.
\newblock


\bibitem[\protect\citeauthoryear{Tsur, Littman, and Rappoport}{Tsur
  et~al\mbox{.}}{2013}]%
        {tsur2013efficient}
\bibfield{author}{\bibinfo{person}{Oren Tsur}, \bibinfo{person}{Adi Littman},
  {and} \bibinfo{person}{Ari Rappoport}.} \bibinfo{year}{2013}\natexlab{}.
\newblock \showarticletitle{Efficient Clustering of Short Messages into General
  Domains.}. In \bibinfo{booktitle}{{\em ICWSM}}. Citeseer.
\newblock


\bibitem[\protect\citeauthoryear{Twitter}{Twitter}{2009}]%
        {twitterspam}
\bibfield{author}{\bibinfo{person}{Twitter}.} \bibinfo{year}{2009}\natexlab{}.
\newblock \bibinfo{title}{Reporting spam on Twitter}.
\newblock
  \bibinfo{howpublished}{\url{https://support.twitter.com/articles/64986}}.
  (\bibinfo{year}{2009}).
\newblock


\bibitem[\protect\citeauthoryear{Twitter}{Twitter}{2015}]%
        {twittercomp}
\bibfield{author}{\bibinfo{person}{Twitter}.} \bibinfo{year}{2015}\natexlab{}.
\newblock \bibinfo{title}{Twitter usage}.
\newblock \bibinfo{howpublished}{\url{https://about.twitter.com/company}}.
  (\bibinfo{year}{2015}).
\newblock


\bibitem[\protect\citeauthoryear{Viswanath, Bashir, Crovella, Guha, Gummadi,
  Krishnamurthy, and Mislove}{Viswanath et~al\mbox{.}}{2014}]%
        {viswanath2014towards}
\bibfield{author}{\bibinfo{person}{Bimal Viswanath},
  \bibinfo{person}{Muhammad~Ahmad Bashir}, \bibinfo{person}{Mark Crovella},
  \bibinfo{person}{Saikat Guha}, \bibinfo{person}{Krishna~P Gummadi},
  \bibinfo{person}{Balachander Krishnamurthy}, {and} \bibinfo{person}{Alan
  Mislove}.} \bibinfo{year}{2014}\natexlab{}.
\newblock \showarticletitle{Towards Detecting Anomalous User Behavior in Online
  Social Networks.}. In \bibinfo{booktitle}{{\em Usenix Security}},
  Vol.~\bibinfo{volume}{14}.
\newblock


\bibitem[\protect\citeauthoryear{Wang}{Wang}{2010}]%
        {wang2010detecting}
\bibfield{author}{\bibinfo{person}{Alex~Hai Wang}.}
  \bibinfo{year}{2010}\natexlab{}.
\newblock \showarticletitle{Detecting spam bots in online social networking
  sites: a machine learning approach}.
\newblock In \bibinfo{booktitle}{{\em Data and Applications Security and
  Privacy XXIV}}. \bibinfo{publisher}{Springer}, \bibinfo{pages}{335--342}.
\newblock


\bibitem[\protect\citeauthoryear{Wang, Konolige, Wilson, Wang, Zheng, and
  Zhao}{Wang et~al\mbox{.}}{2013}]%
        {wang2013you}
\bibfield{author}{\bibinfo{person}{Gang Wang}, \bibinfo{person}{Tristan
  Konolige}, \bibinfo{person}{Christo Wilson}, \bibinfo{person}{Xiao Wang},
  \bibinfo{person}{Haitao Zheng}, {and} \bibinfo{person}{Ben~Y Zhao}.}
  \bibinfo{year}{2013}\natexlab{}.
\newblock \showarticletitle{You Are How You Click: Clickstream Analysis for
  Sybil Detection.}. In \bibinfo{booktitle}{{\em Usenix Security}},
  Vol.~\bibinfo{volume}{14}.
\newblock


\bibitem[\protect\citeauthoryear{Wang, Bai, Stanton, Chen, and Chang}{Wang
  et~al\mbox{.}}{2009}]%
        {wang2009plda}
\bibfield{author}{\bibinfo{person}{Yi Wang}, \bibinfo{person}{Hongjie Bai},
  \bibinfo{person}{Matt Stanton}, \bibinfo{person}{Wen-Yen Chen}, {and}
  \bibinfo{person}{Edward~Y Chang}.} \bibinfo{year}{2009}\natexlab{}.
\newblock \showarticletitle{Plda: Parallel latent dirichlet allocation for
  large-scale applications}. In \bibinfo{booktitle}{{\em International
  Conference on Algorithmic Applications in Management}}. Springer,
  \bibinfo{pages}{301--314}.
\newblock


\bibitem[\protect\citeauthoryear{Weng, Lim, Jiang, and He}{Weng
  et~al\mbox{.}}{2010}]%
        {weng2010twitterrank}
\bibfield{author}{\bibinfo{person}{Jianshu Weng}, \bibinfo{person}{Ee-Peng
  Lim}, \bibinfo{person}{Jing Jiang}, {and} \bibinfo{person}{Qi He}.}
  \bibinfo{year}{2010}\natexlab{}.
\newblock \showarticletitle{Twitterrank: finding topic-sensitive influential
  twitterers}. In \bibinfo{booktitle}{{\em Proceedings of the third ACM
  international conference on Web search and data mining}}. ACM,
  \bibinfo{pages}{261--270}.
\newblock


\bibitem[\protect\citeauthoryear{Weng, Menczer, and Ahn}{Weng
  et~al\mbox{.}}{2013}]%
        {weng2013virality}
\bibfield{author}{\bibinfo{person}{Lilian Weng}, \bibinfo{person}{Filippo
  Menczer}, {and} \bibinfo{person}{Yong-Yeol Ahn}.}
  \bibinfo{year}{2013}\natexlab{}.
\newblock \showarticletitle{Virality prediction and community structure in
  social networks}.
\newblock \bibinfo{journal}{{\em Scientific reports\/}}  \bibinfo{volume}{3}
  (\bibinfo{year}{2013}).
\newblock


\bibitem[\protect\citeauthoryear{Wu, Goel, and Davison}{Wu
  et~al\mbox{.}}{2006}]%
        {wu2006topical}
\bibfield{author}{\bibinfo{person}{Baoning Wu}, \bibinfo{person}{Vinay Goel},
  {and} \bibinfo{person}{Brian~D Davison}.} \bibinfo{year}{2006}\natexlab{}.
\newblock \showarticletitle{Topical trustrank: Using topicality to combat web
  spam}. In \bibinfo{booktitle}{{\em Proceedings of the 15th international
  conference on World Wide Web}}. ACM, \bibinfo{pages}{63--72}.
\newblock


\bibitem[\protect\citeauthoryear{Xu, Zhang, and Zhu}{Xu et~al\mbox{.}}{2010}]%
        {xu2010toward}
\bibfield{author}{\bibinfo{person}{Wei Xu}, \bibinfo{person}{Fangfang Zhang},
  {and} \bibinfo{person}{Sencun Zhu}.} \bibinfo{year}{2010}\natexlab{}.
\newblock \showarticletitle{Toward worm detection in online social networks}.
  In \bibinfo{booktitle}{{\em Proceedings of the 26th Annual Computer Security
  Applications Conference}}. ACM, \bibinfo{pages}{11--20}.
\newblock


\bibitem[\protect\citeauthoryear{Yang, Harkreader, and Gu}{Yang
  et~al\mbox{.}}{2011}]%
        {yang2011free}
\bibfield{author}{\bibinfo{person}{Chao Yang}, \bibinfo{person}{Robert~Chandler
  Harkreader}, {and} \bibinfo{person}{Guofei Gu}.}
  \bibinfo{year}{2011}\natexlab{}.
\newblock \showarticletitle{Die free or live hard? empirical evaluation and new
  design for fighting evolving twitter spammers}. In \bibinfo{booktitle}{{\em
  Recent Advances in Intrusion Detection (RAID)}}.
\newblock


\bibitem[\protect\citeauthoryear{Yardi, Romero, Schoenebeck,
  et~al\mbox{.}}{Yardi et~al\mbox{.}}{2009}]%
        {yardi2009detecting}
\bibfield{author}{\bibinfo{person}{Sarita Yardi}, \bibinfo{person}{Daniel
  Romero}, \bibinfo{person}{Grant Schoenebeck}, {et~al\mbox{.}}}
  \bibinfo{year}{2009}\natexlab{}.
\newblock \showarticletitle{Detecting spam in a twitter network}.
\newblock \bibinfo{journal}{{\em First Monday\/}} \bibinfo{volume}{15},
  \bibinfo{number}{1} (\bibinfo{year}{2009}).
\newblock


\bibitem[\protect\citeauthoryear{Ye and Wu}{Ye and Wu}{2010}]%
        {ye2010measuring}
\bibfield{author}{\bibinfo{person}{Shaozhi Ye} {and} \bibinfo{person}{S~Felix
  Wu}.} \bibinfo{year}{2010}\natexlab{}.
\newblock \bibinfo{booktitle}{{\em Measuring message propagation and social
  influence on Twitter}}.
\newblock \bibinfo{publisher}{Springer}.
\newblock


\end{thebibliography}
